\documentclass[%
 reprint,
unsortedaddress,
 amsmath,amssymb,
prb,
floatfix,
showpacs,superscriptaddress]{revtex4-1}

\usepackage{graphicx}
\usepackage{dcolumn}
\usepackage{bm}

\begin{document}


\title{Xe Irradiation of Graphene on Ir(111): From Trapping to Blistering}

\author{Charlotte Herbig}
\email{herbig@ph2.uni-koeln.de}
\affiliation{II. Physikalisches Institut, Universit{\"a}t zu K{\"o}ln, Z{\"u}lpicher Stra{\ss}e 77, 50937 K{\"o}ln, Germany}
\author{E. Harriet {\AA}hlgren}
\affiliation{Department of Physics, University of Helsinki, P.O. Box 64, FI-00014 Helsinki, Finland}
\author{Ulrike A. Schr{\"o}der}
\affiliation{II. Physikalisches Institut, Universit{\"a}t zu K{\"o}ln, Z{\"u}lpicher Stra{\ss}e 77, 50937 K{\"o}ln, Germany}
\author{Antonio J. Mart\'{i}nez-Galera}
\affiliation{II. Physikalisches Institut, Universit{\"a}t zu K{\"o}ln, Z{\"u}lpicher Stra{\ss}e 77, 50937 K{\"o}ln, Germany}
\author{Mohammad A. Arman}
\affiliation{Division of Synchrotron Radiation Research, Lund University, Box 118, 22100 Lund, Sweden}
\author{Jani Kotakoski}
\affiliation{Faculty of Physics, University of Vienna, Boltzmanngasse 5, 1090 Wien, Austria}
\author{Jan Knudsen}
\affiliation{Max IV Laboratory and Division of Synchrotron Radiation Research, Lund University, Box 118, 22100 Lund, Sweden}
\author{Arkady V. Krasheninnikov}
\affiliation{Department of Applied Physics, Aalto University, P.O. Box 11100, FI-00076 Aalto, Finland}
\author{Thomas Michely}
\affiliation{II. Physikalisches Institut, Universit{\"a}t zu K{\"o}ln, Z{\"u}lpicher Stra{\ss}e 77, 50937 K{\"o}ln, Germany}

\date{\today}

\begin{abstract}
Using X-ray photoelectron spectroscopy, thermal desorption spectroscopy, and scanning tunneling microscopy we show that upon keV Xe$^+$ irradiation of graphene on Ir(111), Xe atoms are trapped under the graphene. Upon annealing, aggregation of Xe leads to graphene bulges and blisters. The efficient trapping is an unexpected and remarkable phenomenon, given the absence of chemical binding of Xe to Ir and to graphene, the weak interaction of a perfect graphene layer with Ir(111), as well as the substantial damage to graphene due to irradiation. By combining molecular dynamics simulations and density functional theory calculations with our experiments, we uncover the mechanism of trapping. We describe ways to avoid blister formation during graphene growth, and also demonstrate how ion implantation can be used to intentionally create blisters without introducing damage to the graphene layer. Our approach may provide a pathway to synthesize new materials at a substrate - 2D material interface or to enable confined reactions at high pressures and temperatures.

\end{abstract}

\maketitle


\section{Introduction}

When graphene (Gr) adheres to a substrate, Gr blisters or bubbles can be formed by intentionally pressurizing a cavity in the substrate \cite{Bunch2008,Koenig2011,Zabel2012,Koenig2012,Boddeti2013}, by agglomeration of atoms or molecules residing or being created in the space between Gr and the substrate \cite{Stolyarova2009,Georgiou2011,Lim2013}, or even as a result of thermal processing \cite{Pan2012}. Such bubbles and blisters in Gr are of considerable interest in view of potential applications ranging from microscopic adaptive-focus lenses \cite{Georgiou2011} via fluid cells for transmission electron microscopy imaging \cite{Yuk2012} or high pressure/~high temperature chemistry \cite{Lim2013} to surfaces with properties switchable by pressure \cite{Boddeti2013}. Blister and bubble formation have also received considerable interest from a more fundamental point of view, since the strain associated with them gives rise to huge pseudomagnetic fields in Gr \cite{Guinea2009,Levy2010,Lu2012} and thus to a quantization of the electronic structure into Landau levels. 

Bubble formation, blistering, and exfoliation in consequence of implantation of light elements like He, D, or H are phenomena that have been thoroughly investigated in the context of nuclear and fusion materials \cite{Scherzer1983}. These phenomena were turned into useful technology by Bruel when applied to achieve microslicing of Si-wafers in silicon on insulator technology \cite{Bruel1995}. Similarly, He$^+$ irradiation of graphite in the basal plane orientation with a few ten keV ion energy has been found to result in gas bubble formation, blistering, and exfoliation \cite{Alimov1995,Chernikov1996}. 
Using low energy B and N ion implantation, substitutional doping of freestanding Gr was pioneered by Bangert \textit{et al.} \cite{Bangert2013}. With similar low ion energies substitutional N implantation of Gr on SiC was accomplished by Telychko \textit{et al.} \cite{Telychko2014}. For low energies and fluences (negligible damage), Cun \textit{et al.} \cite{Cun2013,Cun2014a,Cun2014b,Cun2015} demonstrated that implanted Ar$^+$ ions may come to rest under a 2D-layer strongly adhering to a metal surface and remain there upon annealing. In a recent comment \cite{Herbig2015} to a paper by Herbig \textit{et al.} \cite{Herbig2014}, we pointed out that Xe$^+$ irradiation of Gr on Ir(111) is accompanied by Xe trapping at the interface. 

Here we demonstrate that even a defective Gr layer on Ir(111) effectively traps Xe supplied by ion irradiation up to very high temperatures. Upon annealing the trapped Xe aggregates cause flat bulges in Gr that transform eventually into highly pressurized blisters. Using a combined experimental and theoretical approach, we uncover that efficient Xe trapping is favored by collisional (reduced ion reflection due to Gr cover) and chemical (strong Gr edge bonds with Ir) effects. By proper noble gas pre-implantation and thermal processing, we grow noble gas filled Gr blisters with tunable size, areal density, and without introducing ion irradiation damage. The observed phenomena are expected to take place for a variety of substrates, ion species, and 2D-layers. 
We envision that implantation of atoms with low solubility into a substrate covered with Gr (or another 2D-layer) followed by annealing could be used for stimulating confined reactions in the space between Gr and the substrate, and as 2D-layers are transparent to light, for confined photo-chemical reactions. Such an approach could also be used for reactive growth of interfacial layers in between the substrate and a 2D-layer.

\section{Methods}

The experiments were conducted in ultra high vacuum setups in Cologne [scanning tunneling microscopy (STM), thermal desorption spectroscopy (TDS)] and at the I311 beamline of Max IV Laboratory in Lund [X-ray photoelectron spectroscopy (XPS)]. Cleaning of Ir(111) was conducted by cycles of Xe$^+$ (Cologne) or Ar$^+$ (Lund), irradiation at room temperature (r.t.), O$_2$ treatment at 1120\,K and annealing to 1520\,K (Cologne) or 1300\,K (Lund). A well-oriented and closed Gr layer was grown by one cycle of r.t. C$_2$H$_4$ adsorption until saturation, subsequent thermal decomposition at 1300\,K followed by Gr growth using 100\,L ethylene at 1170\,K \cite{vanGastel2009}. For the experiments represented by Fig.~\ref{fig:Fig5} and Fig. S6 of the Supplemental Material (SM) \cite{SuppInfo} both temperatures were set to 1200\,K. The ion flux during Xe$^+$ irradiation at 300\,K was controlled through adjusting the sample current that itself was calibrated for each ion energy by a Faraday cup in the Cologne setup. Ion incidence was $\approx 30^\circ$ with respect to the surface normal for the XPS experiments (Lund) and at normal incidence for all other experiments (Cologne). Sputtering yields and ion ranges deviate by less than 10\% for these two angles of incidence. Annealing was conducted for 300\,s at 1000\,K and 120\,s at 1300\,K, if not specified otherwise. The resulting sample morphology was imaged by STM at r.t. with a typical sample bias $U_s\,\approx\,-1\,V$ and tunneling current $I_t\,\approx\,1\,nA$. All XP-spectra were collected at r.t. in normal emission using a photon energy of 1000\,eV for Xe\,3d and Ir\,4s with a total energy resolution of light and analyzer better than 400\,meV. After subtraction of a polynomial background the spectra were normalized to the height of the Ir\,4s peak. TDS was conducted with a heating rate of 5\,K/s.

The classical molecular dynamics (MD) simulations were run with the PARCAS code~\cite{Nordlund1998} including only the atomic interactions, as for low ion energies electronic stopping can be neglected to good approximation. For each energy and system, 96 to 300 ions were shot, each separately, in perpendicular direction towards the system at a randomly chosen impact point. To model the substrate, we used Pt ($Z=78$, mass 195\,u) instead of an Ir substrate ($Z=77$, mass 192\,u) because a well-established interaction model exists for Pt-Pt and Pt-C \cite{Albe2002}, unlike for Ir-Ir and Ir-C. Due to the similarity in atomic masses, structure, and chemistry of these two species, only minor differences are expected in the MD simulation results. The simulation cell size and time were adapted to allow sufficient relaxation of the system prior to analysis. The cell consisted of 18360 to 45900 Pt atoms to which 2584 C atoms were added for Gr/Pt(111). The simulation time ranged from 0.3\,ns to 0.5\,ns. Heat dissipation at the system edges was modeled with the Berendsen thermostat~\cite{Berendsen1984}. The distance between the Gr sheet and the substrate was set to 3.31~{\AA} in agreement with the experimentally determined height of 3.3~{\AA} \cite{Sutter2009}. Before the final relaxation the substrate lattice constant was stretched by 1\,\% in order to coincide the two subsystems. The Pt-Pt and Pt-C interactions were modeled by the potentials of Albe \textit{et al.}~\cite{Albe2002}, those of C-C by the bond-order potential of Brenner \textit{et al.}~\cite{Brenner2002} with a repulsive part~\cite{ZBL} for small atom separations, and the ion interactions by the universal repulsive potential~\cite{ZBL}. 
 
The first-principles density functional theory (DFT) calculations were performed using the
plane-wave-basis-set Vienna \textit{ab initio} simulation package\cite{VASP1,VASP2}. The
projector augmented wave approach \cite{PAW2} was used to describe the core
electrons and a non-local van der Waals functional \cite{Bjorkman2012} to
describe exchange and correlation.  A plane wave kinetic energy cutoff of
400\,eV was found to converge energy differences between different
configurations within 0.1 eV. The same accuracy was achieved with regard to the
number of k-points ($3\times 3 \times1 $) in the two-dimensional Brillouin
zone. All structures were relaxed until atomic forces were below 0.002 eV/\AA.
The calculations were carried out for a 200-atom $10 \times 10$ Gr supercell on
top of a $9\times9$ three-atomic-layer-thick Ir(111) slab containing 243 atoms,
as in our previous work \cite{Standop2013,Blanc2013}, or for the same Ir slab
without Gr. 20~{\AA} of vacuum was added in the
transverse direction to separate the periodic images of the slab.
The simulation supercell used for the calculations of the migration 
path/energy of Xe from under graphene is shown in SM \cite{SuppInfo}. In practice, the Xe atom was 
placed in several positions along the path, and the atom was allowed to move 
only perpendicular to Ir(111) during geometry optimization. 
All other atoms were allowed to move without any constraints.

\section{Results and Discussion}

\begin{figure}[htbp]
		\includegraphics{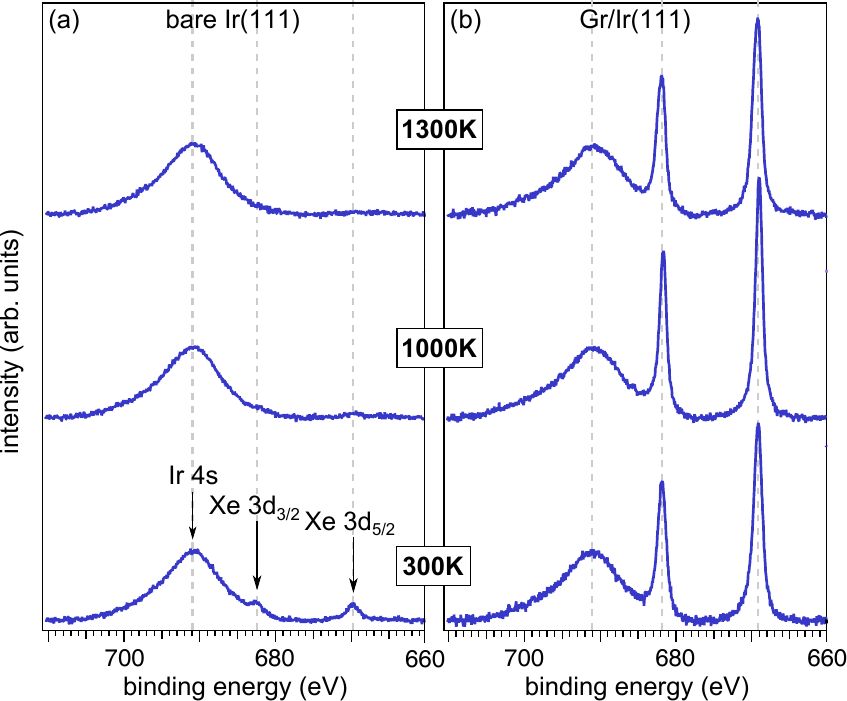}
		\caption{(color online) XP-spectra of the Ir\,4s, Xe\,3d$_{3/2}$, and Xe\,3d$_{5/2}$ core levels for (a) bare Ir(111) and (b) Gr covered Ir(111) after exposure to 0.1\,MLE of 0.5\,keV Xe$^+$ at 300\,K and annealed to stepwise increasing temperatures, as indicated. 
		\label{fig:Fig1}}
\end{figure}

In the first set of XPS experiments we exposed a bare Ir(111) sample to 0.5\,keV Xe$^+$ at 300\,K and conducted successive annealing to 1000\,K and 1300\,K. The ion fluence was 0.1\,MLE, where 1\,MLE is $1.57 \times 10^{19}$\,ions/m$^2$, i.e., numerically identical to the surface atomic density of Ir(111). The bottom XP spectrum in Fig.~\ref{fig:Fig1}(a) displays the Ir\,4s core level peak together with the Xe\,3d$_{3/2}$/Xe\,3d$_{5/2}$ doublet after irradiation at 300\,K. The Xe\,3d$_{5/2}$ binding energy (BE) of 669.7\,eV is consistent with values measured for implanted or adsorbed Xe at other metals \cite{Baba1993,Behm1986} \footnote{For Gr/Ir(111) the Xe\,3d$_{5/2}$ peak energy is between 669.0\,eV and 669.2\,eV at all temperatures, while for bare Ir(111) at 300\,K it is 669.7\,eV. This variation is presumably due to a better extra-atomic screening of the implanted Xe as opposed to the trapped one \cite{Baba1993}, but its explanation is beyond the scope of the present investigation.}. Since Xe does not adsorb on Ir(111) at 300\,K (see our calculations below and ref. \onlinecite{Widdra1998}), the Xe~3d signal must be attributed to Xe implanted into the Ir sample. Upon annealing to 1000\,K the Xe~3d signal diminishes and has vanished after annealing to 1300\,K [middle and top spectra in Fig.~\ref{fig:Fig1}(a)]. Consistent with similar studies for noble gas irradiation of other metals \cite{Donnelly1978,Lahrood2011}, we explain these changes as follows: due to thermal excitation Xe is released from its capture sites inside the crystal, diffuses to the surface, and finally desorbs to the vacuum, such that only little (1000\,K) or no (1300\,K) Xe is left in the crystal.
 
Conducting precisely the same irradiation experiment, but for Ir(111) fully covered by Gr [hereafter named Gr/Ir(111)], yields dramatically different XP spectra as shown in Fig.~\ref{fig:Fig1}(b). Already after irradiation at 300\,K [bottom spectrum of Fig.~\ref{fig:Fig1}(b)] the integrated Xe\,3d$_{5/2}$ intensity is an order of magnitude higher (compare Table~\ref{tab:table1}). All the additional intensity (and thus nearly all intensity) must be due to Xe trapped in between the Gr cover and the Ir substrate. Annealing to 1000\,K and 1300\,K hardly changes the integrated Xe\,3d peak intensity. The interpretation is straightforward: trapped Xe did not escape to any significant extent from under Gr to the vacuum, despite the fact that the Gr cover was heavily damaged by the Xe$^+$ irradiation. 

Table~\ref{tab:table1} compiles the integrated Xe\,3d$_{5/2}$ intensities after ion exposure and annealing for bare Ir(111) and Gr/Ir(111). The intensities are given in \% of the Xe\,3d$_{5/2}$ intensity of an adsorbed, saturated Xe layer on bare Ir(111). Assuming a Xe saturation coverage of 0.33\,ML on Ir(111) \cite{Kern1988} and neglecting attenuation effects, division of the tabulated intensities by a factor of three yields the amounts of trapped Xe in ML with respect to Ir(111). Thus, after annealing to 1000\,K, 4.1\,\%ML Xe is estimated to be trapped, i.e., more than 40\,\% of the incident ions. This value is a lower bound, since the Ir~4s photoelectron attenuation through the adsorbed Xe layer for Xe/Ir(111) and the attenuation of the Xe~3d signal from deeper layers in Ir are neglected. Both factors tend to increase our estimate. Irrespective of these uncertainties, it is obvious that a substantial fraction of the incident Xe remains underneath Gr up to the highest annealing temperature of 1300\,K.
 
\begin{table}[htbp]
\caption{Integrated Xe\,3d$_{5/2}$ intensities $I_{\rm{bare}}$ and $I_{\rm{Gr}}$ of bare and Gr covered Ir(111), respectively, after exposure to 0.1\,MLE 0.5\,keV or 3\,keV Xe$^+$ at 300\,K as well as annealing to  1000\,K and 1300\,K. Also, the intensity ratio $I_{\rm{Gr}} / I_{\rm{bare}}$ is specified for the ion energies and temperatures used. Intensities are given as \% of the integrated Xe\,3d$_{5/2}$ intensity of a saturated Xe layer on Ir(111) and are calibrated to the Ir~4s peak height measured simultaneously.  
\label{tab:table1}
}
\begin{ruledtabular}
\begin{tabular}{lcccccc}

                                    & \multicolumn{3}{c}{0.5\,keV}             & \multicolumn{3}{c}{3\,keV}                   \\ \colrule
                                    & \textrm{$I_{\rm{bare}}$} & \textrm{$I_{\rm{Gr}}$} &\textrm{$I_{\rm{Gr}} / I_{\rm{bare}}$} &\textrm{$I_{\rm{bare}}$} & \textrm{$I_{\rm{Gr}}$} &\textrm{$I_{\rm{Gr}} / I_{\rm{bare}}$}\\ \colrule
\textrm{300\,K}                 &     1.3  & 11.5  & 9  & 2.9 & 5.9  & 2     \\
\textrm{1000\,K}                &     0.4  & 12.2  & 34 & 1.9 & 6.0  & 3    \\
\textrm{1300\,K}                &    $< 0.2$  & 11.6  & $> 60$ & 1.5 & 8.0  & 6    \\
\end{tabular}
\end{ruledtabular}
\end{table}

\begin{figure}[htbp]
		\includegraphics{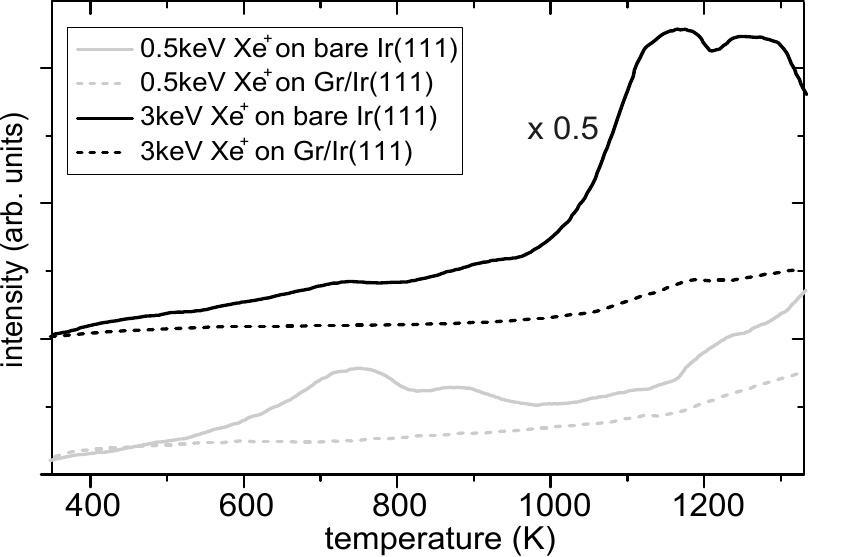}
		\caption{TDS of Xe (131 amu) after exposure of bare Ir(111) and Gr/Ir(111) to 0.1 MLE Xe$^+$ at 300\,K. Solid gray line: 0.5\,keV, bare Ir(111); dashed gray line: 0.5\,keV, Gr/Ir(111); solid black line: 3\,keV, bare Ir(111); dashed black line: 3\,keV, Gr/Ir(111). Heating rate 5\,K/s. For clarity, the 3\,keV data is shifted vertically and multiplied by a factor of 0.5.
		\label{fig:Fig2}}
\end{figure}

While XPS tells about the Xe remaining at the sample after irradiation, a complementary view is provided by TDS, which yields information about the Xe leaving the sample after irradiation through thermal activation. The solid red line in Fig.~\ref{fig:Fig2} is the thermal desorption trace of Xe after 0.5\,keV irradiation of bare Ir(111) with similar parameters as used for the XPS experiment. It exhibits a double peak structure with peak desorption temperatures of 745\,K and 880\,K. Beyond 1000\,K the desorption rate gradually increases up to the end of the heating ramp at 1330\,K. The dashed red line in Fig.~\ref{fig:Fig2} is the thermal desorption trace of Xe after irradiation of Gr/Ir(111). It is a flat featureless curve with its slope increasing slightly for temperatures beyond 1000\,K. Despite an order of magnitude more Xe in the Gr/Ir(111) sample, the integrated amount of Xe desorbing is lower by about a factor of two.
Therefore, our TDS data elucidate that only a small fraction of the total amount of Xe in the Gr/Ir(111) sample desorbs up to 1300\,K. 

The experimental data are substantiated by our MD simulations (compare Table~\ref{tab:table2}), for which we used a Pt instead of an Ir substrate (see Methods section). For 0.5\,keV Xe hitting bare Pt(111) at normal incidence, when the energetic phase of the collision has ended (0.5\,ns after the impact), 98\,\% of the impinging Xe has been reflected (i.e., returned to vacuum) and only 2\,\% of the primary ions have been implanted into the Pt crystal (into a depth of less than three atomic layers). The same simulations with Xe hitting Gr/Pt(111) yield a strikingly different picture. Then, only 2\,\% of the projectiles are reflected and 5\,\% are implanted into Pt, but 93\,\% of the Xe ions are stuck in between Gr and the Pt(111) surface [see Fig. S1 of the SM for depth distributions \cite{SuppInfo}]. Apparently, the primary Xe ions penetrate the Gr sheet easily; much more easily than a Pt layer. The reason for this is twofold: (i) due to the much smaller nuclear charge of C ($Z=12$) compared to Pt ($Z=78$) the scattering cross section of Xe with C is much smaller than that with Pt (and Ir); (ii) even in a head-on collision with a C atom the Xe atom loses less than a third of its energy and continues to move towards the metallic substrate. There, the Xe projectiles effectively transfer energy to the substrate Pt atoms (mass ratio 131\,u to 195\,u), such that in case of momentum reversal they no longer possess enough energy to pass through the Gr sheet and are instead trapped. Note also that when impinging the Gr sheet from below, the Xe projectiles have not only much less energy and thus a larger scattering cross section with C, but under nearly all circumstances off-normal or grazing direction, such that the transparency of the Gr sheet for ion passage is strongly reduced. The results of the MD simulations are fully consistent with the XPS observation that an order of magnitude more Xe remains within the sample if there is a Gr cover present (compare Fig.~\ref{fig:Fig1} and Table~\ref{tab:table1}). Still, they do not explain why the trapped Xe does not escape to the vacuum by diffusion through irradiation induced vacancies and vacancy islands in the Gr sheet. Note also that the MD simulations are conducted for a perfect Gr sheet, while in the experiment ion induced damage of successive impacts accumulates. 

\begin{table}[htbp]
\caption{Fraction of reflected, trapped, or implanted primary Xe ions given in \% as obtained by MD simulations $ > 95$ events per case. Also specified is the average depth of implantation $\overline{r}_{im}$ measured from the topmost Pt layer. 
\label{tab:table2}}
\begin{ruledtabular}
\begin{tabular}{lcccc}
                                   & \multicolumn{2}{c}{0.5\,keV}             & \multicolumn{2}{c}{3\,keV}                   \\ \colrule 
                                   & \textrm{Pt(111)} & \textrm{Gr/Pt(111)} & \textrm{Pt(111)} & \textrm{Gr/Pt(111)} \\ \colrule 
\textrm{reflected}                 &    98              &  2                     &  57                    &   6                    \\
\textrm{trapped}                   &    -              &  93                     &  -                    &   41                    \\
\textrm{implanted}                 &    2              &  5                     &  43                    &   53                    \\
$\overline{r}_{im}$ (nm)          &    0.4              &  0.2                     &  1                    &   0.9              \\
\end{tabular}
\end{ruledtabular}
\end{table}

\begin{figure}[t]
		\includegraphics{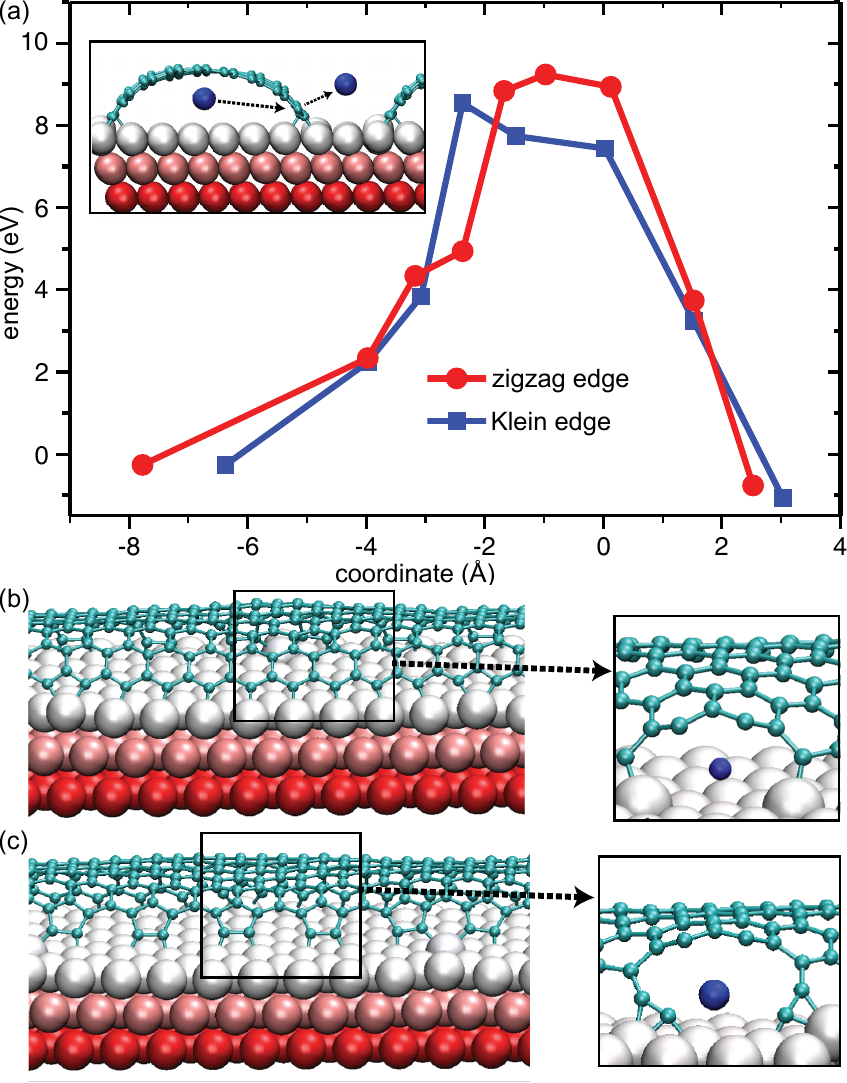}
		\caption{(color online) (a) System energy as a function of the reaction coordinate (0 corresponds to the Xe atom at the Gr edge position) for two different edge structures and the path of a Xe atom schematically depicted in the inset. (b) Structure of zig-zag edge (left) and zoomed structure with the Xe atom in the transition state (right). (c) Structure of the reconstructed Klein edge (left) and zoomed structure with the Xe atom in the transition state (right).  
		\label{fig:Fig3}}
\end{figure}

To understand why trapped Xe remains under a defective Gr layer, we carried out DFT calculations, as described in the Methods section. We first evaluated the adsorption energy of Xe atoms on bare Ir(111) and Gr/Ir(111) to be $-0.21$\,eV and $-0.17$\,eV, respectively. These numbers are consistent with previous experimental results for Xe adsorption on Pt(111) ($-0.27$\,eV) \cite{Widdra1998} and Xe adsorption on graphite ($-0.25$\,eV). The adsorption energies also agree up to a factor of about two with more accurate calculation schemes using many-body perturbation theory for Xe adsorbed to graphite, to Gr, or to Ni(111), see e.g. \onlinecite{DaSilva2007, Sheng2010, Silvestrelli2015}. In any case, the DFT numbers calculated here indicate the absence of chemical bonding and are in magnitude more than a factor of 30 smaller than the barriers for the penetration of Xe atoms through the graphene edge, as shown below. Next we inserted a single Xe atom between the Gr sheet and Ir(111). After full geometry optimization, taking the energies of isolated Gr/Ir(111) and the Xe atom as the reference, we found that it costs 2.9\,eV to place a Xe atom between Gr and Ir (see Fig. S2 of the SM for geometry \cite{SuppInfo}). This energy penalty results from the deformation caused by the Xe atom in the Gr sheet (elastic energy) and the local delamination of the Gr sheet from its substrate (delamination energy). Note that escape of a trapped Xe into the Ir bulk is impossible due to the very high interstitial formation energy exceeding 10\,eV. The most likely escape path for Xe is thus to go via Gr edges onto the free Ir terrace, where the Xe atom would desorb rapidly under all temperatures of concern, as schematically depicted in the inset of Fig.~\ref{fig:Fig3}(a). Due to Gr sputtering, such edges are formed in large number during irradiation and annealing, but Gr dangling bonds are saturated due to the strong interaction of Gr with the metal giving rise to bending of the sheet towards the metal surface already at room temperature\cite{Standop2013, Herbig2014}. The energy of the system as a function of the coordinate of the Xe atom along the path is presented in Fig.~\ref{fig:Fig3}(a) for two particular edge structures. First, the zig-zag edge that terminates vacancy islands in the Gr sheet [compare Fig.~S5\,(c) and caption]. Additionally, the reconstructed Klein edge is considered here, as such edge structures develop for freestanding Gr under sputtering conditions \cite{He2014}. For both structures an activation energy of more than 8\,eV is obtained. 
Fig.~\ref{fig:Fig3}(b) and (c) show the atomic structure of Gr edges binding to Ir(111), and the close-ups present the Gr edge structure and the Xe atom in the transition state. For passage through the zig-zag edge [Fig.~\ref{fig:Fig3}(b)] 2 C-Ir bonds must be broken temporarily, while for passing through the reconstructed Klein edge [Fig. 3(c)] two C-C bonds need to be broken in the C pentagons bound to the substrate.  
We also calculated the barrier for escape through an open corner in a Gr vacancy island edge and obtain a similar energy barrier (see Fig. S3 of the SM \cite{SuppInfo}). For an efficient release of Xe at 1300\,K a much lower barrier, below 4\,eV, would be necessary. 
Although computational cost prevents us from exploring the entire space of possible edge structures that may arise during irradiation and annealing, it is clear from our calculations that the passage of a Xe atom from under Gr involves breaking of strong Gr-Ir or Gr-Gr bonds. 
Hence, very little Xe is released even through a highly defective Gr sheet with a large number of smaller and larger vacancy islands as well as for temperatures up to 1300\,K. As the escape of Xe from under Gr is impaired by Gr edge binding, the fraction of trapped Xe calculated in the MD simulations is indeed a good first estimate for the experimentally observed trapping efficiency.

\begin{figure*}[htbp]
		\includegraphics{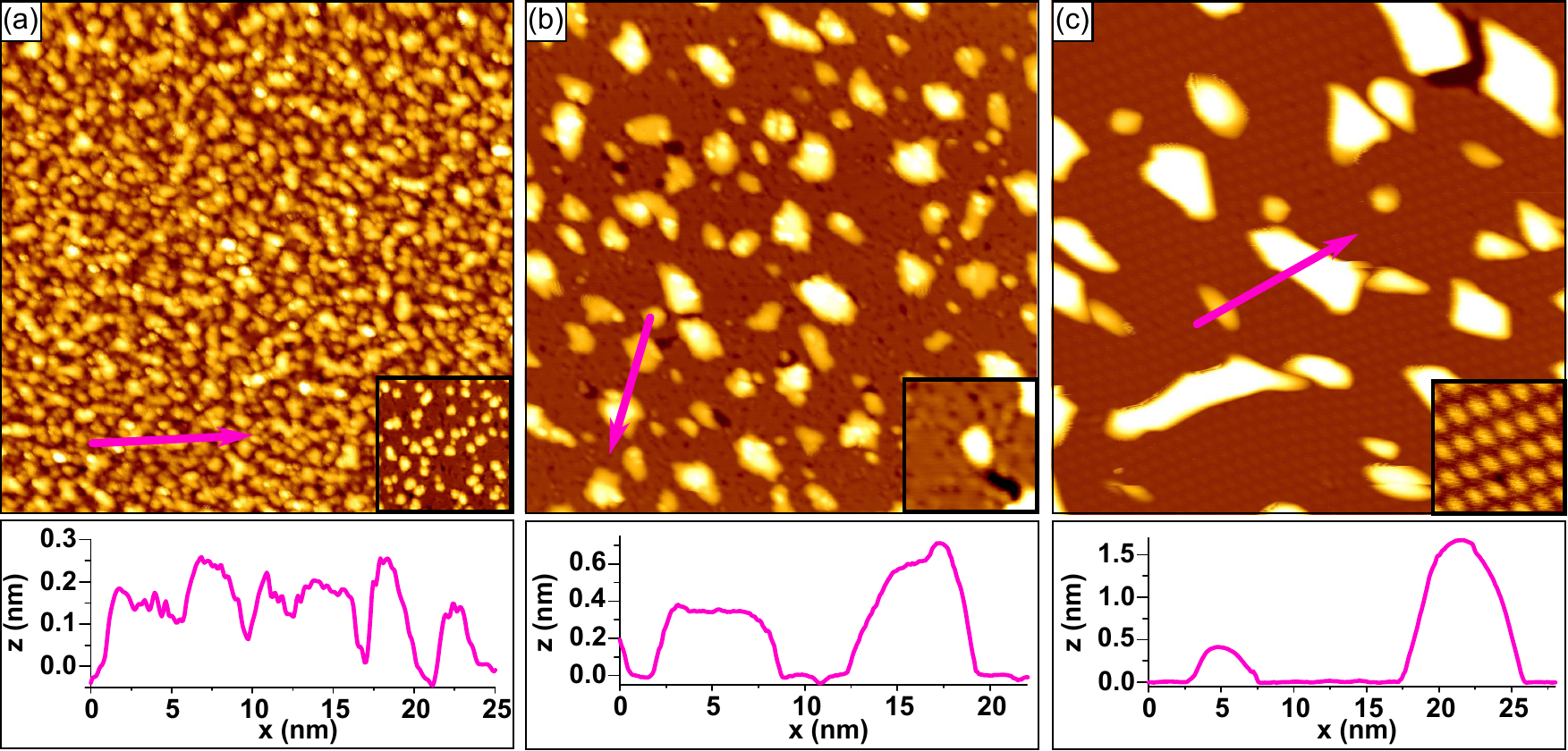}
		\caption{(color online) STM topographs of Gr/Ir(111) and associated height profiles after exposure to 0.1\,MLE of 0.5\,keV Xe$^+$ (a) at 300\,K, (b) annealed to 1000\,K, and (c) annealed to 1300\,K. The height profiles below the topographs are taken along the lines indicated. Insets: (a) bare Ir(111) after exposure to ion irradiation with the same parameters, (b), (c) are zooms of the corresponding topograph. Image sizes are $90\,\rm{nm}\,\times 90\,\rm{nm}$, inset sizes are (a) $24\,\rm{nm}\,\times\,24\,\rm{nm}$ and (b), (c)  $15\,\rm{nm} \times 15\,\rm{nm}$. 
		\label{fig:Fig4}}
\end{figure*}

The real space view of the noble gas trapping under the Gr sheet is presented in the STM topographs of Fig.~\ref{fig:Fig4}. After room temperature irradiation the sample displays a rather dense distribution of small scale protrusions of 0.3\,nm - 0.5\,nm height [compare Fig.~\ref{fig:Fig4}(a)]. Experiments with lower fluence show unambiguously that the protrusions are deformations in the Gr sheet, which itself is defective but globally intact \cite{Herbig2014}. Based on our XPS and TDS experiments, as well as the MD simulations and DFT computations discussed above, we must interpret these protrusions to result to a large extent from Xe atoms, which give rise to Gr's local deformation. Additionally, Ir adatoms pushed onto the Ir(111) surface by collisions \cite{Michely1991} contribute to the formation of protrusions in the Gr sheet. This statement is backed-up by the inset of Fig.~\ref{fig:Fig4}(a), where Ir adatom clusters are visible after irradiation of bare Ir(111) using the same ion beam parameters. 

After annealing the sample to 1000\,K the morphology is dramatically changed [compare Fig.~\ref{fig:Fig4}(b)]. The number of protrusions is strongly reduced and two types of elevations are visible: first, flat bulges with a well defined height around 0.35\,nm, and second, larger, smoothly curved blisters with a maximum height up to 0.8\,nm [height profile in Fig.~\ref{fig:Fig4}(b)]. Also visible are dark areas of few nm size and point-like dark spots with typical separations of 2-3\,nm, as highlighted by the inset. These features are vacancy islands within the Gr sheet, either extended nm-sized areas or small point-like spots pinned to preferential binding sites within the moir\'{e} formed of Gr with the Ir(111) substrate (compare reference \citenum{Standop2013}). The vacancy islands result from aggregation of vacancies created in Gr due to sputtering.  

Upon annealing to 1300\,K the morphology coarsened and is now dominated by few, large blisters with lateral dimensions of the order of 10\,nm and heights around 1.5\,nm, as shown in Fig.~\ref{fig:Fig4}(c). Also some shallower blisters are present. The point-like vacancy clusters have nearly vanished. Large vacancy islands are present in quite low density. One of those is visible in the upper right corner of Fig.~\ref{fig:Fig4}(c), between two large blisters. Apparently the small vacancy clusters detached during annealing from their preferential binding sites and aggregated to large vacancy islands, consistent with the findings of Standop \textit{et al.} \cite{Standop2013}. It is striking that the vacancy island visible in Fig.~\ref{fig:Fig4}(c) is bound by large blisters. This observation underlines the strong binding of Gr to the Ir(111) substrate and the impermeability of the Gr edge for Xe, as found in our DFT calculations. A rather regular moir\'{e} corrugation with a 2.5\,nm pitch indicates that the Gr lattice has restored itself completely [compare inset of Fig.~\ref{fig:Fig4}(c)].

We tentatively interpret the evolution from atomic scale protrusions to large blisters as follows. After room temperature irradiation, a mixture of Xe atoms, Ir adatoms, and detached C atoms is present under the defective Gr cover. Upon annealing to 1000\,K the Ir adatoms and adatom clusters recombine with Ir surface vacancies restoring a flat surface \cite{Standop2013}, while the detached C atoms are largely reincorporated into the Gr sheet. The agglomeration of the remaining Xe atoms (and of the implanted atoms diffusing from the Ir bulk to the surface) gives rise to monolayer Xe platelets that cause flat bulges. Their formation is driven by minimization of deformation and delamination energy within Gr: disperse single Xe atoms cause a much larger deformation of Gr and much more delamination of Gr from the substrate than one flat aggregate containing all Xe atoms. The assumption of a monolayer thick Xe aggregate is consistent with the uniform apparent STM bulge height of about 0.35\,nm, similar to the Xe van der Waals diameter of 0.43\,nm. Since the Gr sheet presses the Xe together, we tentatively assume that also the lateral spacing of the Xe atoms is close to that of solid Xe. This could be realized, e.g., by a ($\sqrt{3} \times \sqrt{3}$)-Xe structure with a coverage of 0.33\,ML \cite{Kern1988}. When the flat bulges grow larger, a shape transformation to smoothly curved blisters takes place. While the delamination energy grows with the square of the linear dimension of the flat bulge, its strain and bending energy - primarily localized along the rim of the bulge - grows only proportional to it. Therefore, at some point a transition to a more three-dimensional shape of the Xe cavity takes place, lowering the delamination energy for a given amount of Xe. After annealing to 1300K the measured blister height and radius (approximating the blister as spherically symmetric) with the calculated adhesion energy of Gr to Ir(111) (50meV per C atom \cite{Busse2011}) allows one to estimate \cite{Koenig2011, Stolyarova2009} pressures of the order of 1GPa for the Xe in the blister. It should be noted that a similar coarsening of gas cavities during annealing, as described here, is also observed for bubbles and blisters after noble gas implantation into bulk materials \cite{Scherzer1983}.    

To explore how trapping depends on ion energy, we conducted the same set of experiments with an ion energy of 3\,keV. Despite the increased Gr sputtering and damaging (the yield for C atom detachment from the Gr sheet increases from 1.0 to 1.8 according to our MD simulations) the general picture is unchanged: the Gr sheet strongly reduces ion reflection and enables efficient Xe trapping upon irradiation at 300\,K. To analyze differences to the 0.5\,keV situation, we first consider changes for the irradiation of bare Ir(111). Compared to 0.5\,keV, for 3\,keV Xe$^+$ into bare metal substantially more Xe is implanted (more than 40\,\% of the impinging Xe atoms) into a larger average depth of 1\,nm, as apparent from the MD results shown in Table~\ref{tab:table2} and consistent with the corresponding XPS data provided as Fig. S4 of the SM \cite{SuppInfo} (compare also ref. \citenum{Herbig2015}). Consequently, substantially more Xe is released from bare Ir during heating as visible in the corresponding Xe desorption trace in Fig.~\ref{fig:Fig2}. However, even after heating to 1300\,K some Xe is left in the Ir crystal, presumably due to incorporation of Xe in stable bulk vacancy clusters formed in consequence of the more violent Xe impacts. With the Gr cover the implantation into Ir does not change much, but trapping under the Gr sheet becomes frequent ($> 40$\,\% of the impinging Xe), thereby strongly suppressing reflection. As for the 0.5\,keV case, the presence of the Gr cover gives rise to a substantial enhancement of the Xe~3d XPS intensity after 300\,K irradiation, which must be attributed to Xe trapping (compare Fig. S4 of the SM \cite{SuppInfo} and Table~\ref{tab:table1}). Xe release is largely prevented by the Gr cover, as obvious from the comparison of bare Ir(111) and Gr/Ir(111) Xe desorption traces after 3\,keV ion irradiation shown in Fig.~\ref{fig:Fig2}. In consequence, by annealing, the amount of trapped Xe continuously increases by Xe streaming to the interface from its implantation sites in the Ir bulk. The STM images display a markedly larger Gr removal resulting from the higher sputtering yield of 3\,keV Xe ions, but otherwise a similar morphological evolution from atomic scale protrusions to large blisters (compare Fig. S5 of the SM \cite{SuppInfo}).

Our experimental analysis, as well as the simulations and calculations presented here indicate that the trapped Xe fully accounts for the bulges and blisters observed. Though we cannot rule out the presence of additional interfacial C, the association of bulge formation with interfacial C (Ref. \citenum{Herbig2014}) appears not to be plausible. More likely, upon annealing the interfacial C reincorporates into the Gr sheet. The trapping yields presented in reference \citenum{Herbig2014} in Figs. 5 and 6 should therefore be interpreted as Xe trapping efficiencies \footnote{To first approximation trapping efficiencies are obtained by dividing the trapping yields specified in reference \citenum{Herbig2014} by a factor of 7.3.}.    
The observation of bulges and blisters after Ar$^+$ and Ne$^+$ irradiation of Gr/Ir(111) in reference \citenum{Herbig2014} indicates efficient trapping under the Gr sheet also for these noble gases. Only for He$^+$ exposure with 0.3\,keV flat bulges and blisters are almost absent, even after an ion fluence of 0.3\,MLE, indicating that trapping does not take place. Presumably, due to the small van der Waals diameter (0.28\,nm for He compared to 0.43\,nm for Xe), He escapes from under Gr without the need to break Gr edge bonds. Indeed, our DFT calculations of He diffusion from under the Gr sheet with the same path as for Xe showed that the barrier for He escape is considerably smaller, on the order of 3-4 eV. Sputtering with He$^+$, preferably even at elevated temperature, may therefore be a method to avoid noble gas trapping during ion irradiation of Gr on a substrate. 

In our view, the phenomenon of trapping after irradiation of 2D-materials is of general relevance. It does not depend sensitively on ion beam parameters (energy, angle of incidence, species, charge state), as long as the ions are able to penetrate the 2D-layer. It should also take place for other combinations of substrate and 2D-layer materials, as long as (i) the implanted species does not react with substrate and 2D-layer, (ii) the solubility of the implanted species in the substrate is low, and (iii) the 2D-layer interacts strongly with its defects and edges to the substrate \cite{Krasheninnikov2014}.

Our investigations indicate that high temperature 2D-layer growth on an improperly degassed substrate with a higher than equilibrium gas concentration may result in inferior quality of the 2D-layer through the formation of gas cavities (blisters, bubbles, wrinkles) at the interface. For example, Gr grown on a metal substrate that was sputter-cleaned with keV noble gas ions and not annealed sufficiently long well \emph{above} the Gr growth temperature, is likely to display gas filled blisters after growth. An example for blister formation during Gr growth on a sputter-cleaned and insufficiently annealed Ir(111) substrate is shown as Fig. S6 of the SM \cite{SuppInfo}. 

\begin{figure*}[htbp]
		\includegraphics[width= 0.75\textwidth]{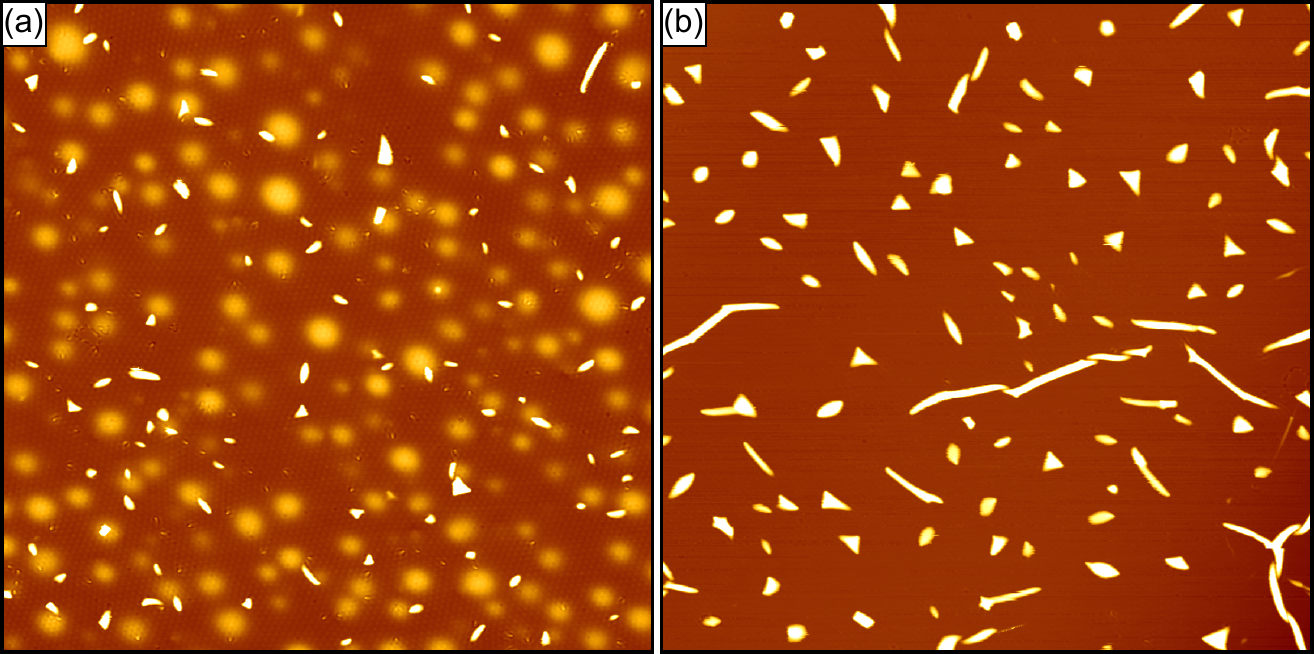}
		\caption{(color online) (a) STM topograph of Gr/Ir(111) after successive exposure to 0.2\,MLE of 5\,keV, 4\,keV, 3\,keV, and 2\,keV Xe$^+$ (accumulated exposure 0.8\,MLE) at 300\,K, subsequent annealing to 1200\,K, and eventually growth of a complete layer of Gr at 1200\,K. (b) sample after additional flash annealing to 1500\,K. Image sizes are 250\,nm $\times$ 250\,nm. 
		\label{fig:Fig5}}
\end{figure*}

In the experiments described up to now, blister formation is a by-product of ion irradiation and always involves damaging of the Gr sheet. However, it is also possible to separate ion irradiation and blister formation, enabling us to create blisters under a perfect Gr sheet with number density and size controllable by thermal processing. To reach this goal we first implanted Xe at 300\,K into bare Ir(111), using sufficiently high and varied ion energies to ensure high gas release temperatures and gas incorporation with a broad depth distribution. To remove surface damage, the sample was flash annealed to 1200\,K and subsequently a Gr sheet was grown at 1200\,K. Fig.~\ref{fig:Fig5}(a) displays the resulting morphology characterized by smooth circular bulges of subatomic height due to gas bubbles in Ir \cite{Michely1991}, as well as irregularly shaped (often triangular) blisters that formed due to gas release from the Ir crystal \emph{during} Gr growth at 1200\,K. The Xe still present in the subsurface bubbles can be released and trapped underneath the perfect Gr sheet by additional annealing to 1500\,K, as shown in Fig.~\ref{fig:Fig5}(b). The bubbles within the Ir crystal vanished, while the volume of the blisters increased by a factor of five. 

We envision that similar procedures of pre-implantation and annealing could be used to create pressurized nanoreactors with one reagent supplied through release of implanted species from the substrate. Also the growth of interfacial layers, e.g., insulators, could be accomplished by providing a reagent from the bulk after 2D-layer growth or transfer and optional intercalation.

\section{Conclusions}

In conclusion, we have determined the mechanisms of atom trapping underneath Gr on Ir(111) during Xe$^+$ ion irradiation. It relies on the combination of two effects. First, for an ion energy up to a few keV a Gr layer on Ir(111) effectively decreases ion reflection in favor of ion trapping in between Gr and its substrate. The efficient trapping results from the easy penetration of Xe through Gr (small scattering cross section) and efficient energy loss to the metallic substrate, making escape after momentum-reversal unlikely. Second, the irradiation induced Gr edges bind strongly to the Ir(111) substrate making escape of Xe through out-diffusion at the Gr edge an unlikely process, even at annealing temperatures of 1300\,K. Driven by elastic and delamination energy of Gr, upon annealing the trapped Xe first forms monolayer thick Xe islands under flat bulges in Gr and, when growing larger at higher temperatures, three-dimensional pressurized blisters in the Gr sheet. The bulge and blister formation upon ion irradiation must be assumed to take place for a broad range of 2D-materials which adhere to the substrate. We demonstrated that ion beam implantation methods in combination with proper thermal processing may be a pathway to enable reactions in pressurized nanocavities or to grow homogeneous interfacial layers between a perfect 2D material and its substrate.

\begin{acknowledgments}
Financial support from the European Commission by a 
Marie Curie Fellowship (A. M.-G.), the Institutional Strategy of the University of Cologne
and the DFG Priority Program 1459 ``Graphene'' within project MI581/20-1 is gratefully acknowledged.
Work at Lund was supported by Swedish research council (grant no. 2012-3850). 
Assistance by the staff at the MAX IV Laboratory is gratefully acknowledged.
The authors thank Carsten Busse and Felix Huttmann for useful discussions.
\end{acknowledgments}



\bibliography{library}

\begin{thebibliography}{54}%
\makeatletter
\providecommand \@ifxundefined [1]{%
 \@ifx{#1\undefined}
}%
\providecommand \@ifnum [1]{%
 \ifnum #1\expandafter \@firstoftwo
 \else \expandafter \@secondoftwo
 \fi
}%
\providecommand \@ifx [1]{%
 \ifx #1\expandafter \@firstoftwo
 \else \expandafter \@secondoftwo
 \fi
}%
\providecommand \natexlab [1]{#1}%
\providecommand \enquote  [1]{``#1''}%
\providecommand \bibnamefont  [1]{#1}%
\providecommand \bibfnamefont [1]{#1}%
\providecommand \citenamefont [1]{#1}%
\providecommand \href@noop [0]{\@secondoftwo}%
\providecommand \href [0]{\begingroup \@sanitize@url \@href}%
\providecommand \@href[1]{\@@startlink{#1}\@@href}%
\providecommand \@@href[1]{\endgroup#1\@@endlink}%
\providecommand \@sanitize@url [0]{\catcode `\\12\catcode `\$12\catcode
  `\&12\catcode `\#12\catcode `\^12\catcode `\_12\catcode `\%12\relax}%
\providecommand \@@startlink[1]{}%
\providecommand \@@endlink[0]{}%
\providecommand \url  [0]{\begingroup\@sanitize@url \@url }%
\providecommand \@url [1]{\endgroup\@href {#1}{\urlprefix }}%
\providecommand \urlprefix  [0]{URL }%
\providecommand \Eprint [0]{\href }%
\providecommand \doibase [0]{http://dx.doi.org/}%
\providecommand \selectlanguage [0]{\@gobble}%
\providecommand \bibinfo  [0]{\@secondoftwo}%
\providecommand \bibfield  [0]{\@secondoftwo}%
\providecommand \translation [1]{[#1]}%
\providecommand \BibitemOpen [0]{}%
\providecommand \bibitemStop [0]{}%
\providecommand \bibitemNoStop [0]{.\EOS\space}%
\providecommand \EOS [0]{\spacefactor3000\relax}%
\providecommand \BibitemShut  [1]{\csname bibitem#1\endcsname}%
\let\auto@bib@innerbib\@empty
\bibitem [{\citenamefont {Bunch}\ \emph {et~al.}(2008)\citenamefont {Bunch},
  \citenamefont {Verbridge}, \citenamefont {Alden}, \citenamefont {van~der
  Zande}, \citenamefont {Parpia}, \citenamefont {Craighead},\ and\
  \citenamefont {McEuen}}]{Bunch2008}%
  \BibitemOpen
  \bibfield  {author} {\bibinfo {author} {\bibfnamefont {J.~S.}\ \bibnamefont
  {Bunch}}, \bibinfo {author} {\bibfnamefont {S.~S.}\ \bibnamefont
  {Verbridge}}, \bibinfo {author} {\bibfnamefont {J.~S.}\ \bibnamefont
  {Alden}}, \bibinfo {author} {\bibfnamefont {A.~M.}\ \bibnamefont {van~der
  Zande}}, \bibinfo {author} {\bibfnamefont {J.~M.}\ \bibnamefont {Parpia}},
  \bibinfo {author} {\bibfnamefont {H.~G.}\ \bibnamefont {Craighead}}, \ and\
  \bibinfo {author} {\bibfnamefont {P.~L.}\ \bibnamefont {McEuen}},\ }\href
  {\doibase 10.1021/nl801457b} {\bibfield  {journal} {\bibinfo  {journal} {Nano
  Lett.}\ }\textbf {\bibinfo {volume} {8}},\ \bibinfo {pages} {2458} (\bibinfo
  {year} {2008})}\BibitemShut {NoStop}%
\bibitem [{\citenamefont {Koenig}\ \emph {et~al.}(2011)\citenamefont {Koenig},
  \citenamefont {Boddeti}, \citenamefont {Dunn},\ and\ \citenamefont
  {Bunch}}]{Koenig2011}%
  \BibitemOpen
  \bibfield  {author} {\bibinfo {author} {\bibfnamefont {S.~P.}\ \bibnamefont
  {Koenig}}, \bibinfo {author} {\bibfnamefont {N.~G.}\ \bibnamefont {Boddeti}},
  \bibinfo {author} {\bibfnamefont {M.~L.}\ \bibnamefont {Dunn}}, \ and\
  \bibinfo {author} {\bibfnamefont {J.~S.}\ \bibnamefont {Bunch}},\ }\href
  {\doibase 10.1038/nnano.2011.123} {\bibfield  {journal} {\bibinfo  {journal}
  {Nat. Nanotechnol.}\ }\textbf {\bibinfo {volume} {6}},\ \bibinfo {pages}
  {543} (\bibinfo {year} {2011})}\BibitemShut {NoStop}%
\bibitem [{\citenamefont {Zabel}\ \emph {et~al.}(2012)\citenamefont {Zabel},
  \citenamefont {Nair}, \citenamefont {Ott}, \citenamefont {Georgiou},
  \citenamefont {Geim}, \citenamefont {Novoselov},\ and\ \citenamefont
  {Casiraghi}}]{Zabel2012}%
  \BibitemOpen
  \bibfield  {author} {\bibinfo {author} {\bibfnamefont {J.}~\bibnamefont
  {Zabel}}, \bibinfo {author} {\bibfnamefont {R.~R.}\ \bibnamefont {Nair}},
  \bibinfo {author} {\bibfnamefont {A.}~\bibnamefont {Ott}}, \bibinfo {author}
  {\bibfnamefont {T.}~\bibnamefont {Georgiou}}, \bibinfo {author}
  {\bibfnamefont {A.~K.}\ \bibnamefont {Geim}}, \bibinfo {author}
  {\bibfnamefont {K.~S.}\ \bibnamefont {Novoselov}}, \ and\ \bibinfo {author}
  {\bibfnamefont {C.}~\bibnamefont {Casiraghi}},\ }\href {\doibase
  10.1021/nl203359n} {\bibfield  {journal} {\bibinfo  {journal} {Nano Lett.}\
  }\textbf {\bibinfo {volume} {12}},\ \bibinfo {pages} {617} (\bibinfo {year}
  {2012})}\BibitemShut {NoStop}%
\bibitem [{\citenamefont {Koenig}\ \emph {et~al.}(2012)\citenamefont {Koenig},
  \citenamefont {Wang}, \citenamefont {Pellegrino},\ and\ \citenamefont
  {Bunch}}]{Koenig2012}%
  \BibitemOpen
  \bibfield  {author} {\bibinfo {author} {\bibfnamefont {S.~P.}\ \bibnamefont
  {Koenig}}, \bibinfo {author} {\bibfnamefont {L.}~\bibnamefont {Wang}},
  \bibinfo {author} {\bibfnamefont {J.}~\bibnamefont {Pellegrino}}, \ and\
  \bibinfo {author} {\bibfnamefont {J.~S.}\ \bibnamefont {Bunch}},\ }\href
  {\doibase 10.1038/nnano.2012.162} {\bibfield  {journal} {\bibinfo  {journal}
  {Nat. Nanotechnol.}\ }\textbf {\bibinfo {volume} {7}},\ \bibinfo {pages}
  {728} (\bibinfo {year} {2012})}\BibitemShut {NoStop}%
\bibitem [{\citenamefont {Boddeti}\ \emph {et~al.}(2013)\citenamefont
  {Boddeti}, \citenamefont {Koenig}, \citenamefont {Long}, \citenamefont
  {Xiao}, \citenamefont {Bunch},\ and\ \citenamefont {Dunn}}]{Boddeti2013}%
  \BibitemOpen
  \bibfield  {author} {\bibinfo {author} {\bibfnamefont {N.~G.}\ \bibnamefont
  {Boddeti}}, \bibinfo {author} {\bibfnamefont {S.~P.}\ \bibnamefont {Koenig}},
  \bibinfo {author} {\bibfnamefont {R.}~\bibnamefont {Long}}, \bibinfo {author}
  {\bibfnamefont {J.}~\bibnamefont {Xiao}}, \bibinfo {author} {\bibfnamefont
  {J.~S.}\ \bibnamefont {Bunch}}, \ and\ \bibinfo {author} {\bibfnamefont
  {M.~L.}\ \bibnamefont {Dunn}},\ }\href {\doibase 10.1115/1.4024255}
  {\bibfield  {journal} {\bibinfo  {journal} {J. Appl. Mech.}\ }\textbf
  {\bibinfo {volume} {80}},\ \bibinfo {pages} {040909} (\bibinfo {year}
  {2013})}\BibitemShut {NoStop}%
\bibitem [{\citenamefont {Stolyarova}\ \emph {et~al.}(2009)\citenamefont
  {Stolyarova}, \citenamefont {Stolyarov}, \citenamefont {Bolotin},
  \citenamefont {Ryu}, \citenamefont {Liu}, \citenamefont {Rim}, \citenamefont
  {Klima}, \citenamefont {Hybertsen}, \citenamefont {Pogorelsky}, \citenamefont
  {Pavlishin}, \citenamefont {Kusche}, \citenamefont {Hone}, \citenamefont
  {Kim}, \citenamefont {Stormer}, \citenamefont {Yakimenko},\ and\
  \citenamefont {Flynn}}]{Stolyarova2009}%
  \BibitemOpen
  \bibfield  {author} {\bibinfo {author} {\bibfnamefont {E.}~\bibnamefont
  {Stolyarova}}, \bibinfo {author} {\bibfnamefont {D.}~\bibnamefont
  {Stolyarov}}, \bibinfo {author} {\bibfnamefont {K.}~\bibnamefont {Bolotin}},
  \bibinfo {author} {\bibfnamefont {S.}~\bibnamefont {Ryu}}, \bibinfo {author}
  {\bibfnamefont {L.}~\bibnamefont {Liu}}, \bibinfo {author} {\bibfnamefont
  {K.~T.}\ \bibnamefont {Rim}}, \bibinfo {author} {\bibfnamefont
  {M.}~\bibnamefont {Klima}}, \bibinfo {author} {\bibfnamefont
  {M.}~\bibnamefont {Hybertsen}}, \bibinfo {author} {\bibfnamefont
  {I.}~\bibnamefont {Pogorelsky}}, \bibinfo {author} {\bibfnamefont
  {I.}~\bibnamefont {Pavlishin}}, \bibinfo {author} {\bibfnamefont
  {K.}~\bibnamefont {Kusche}}, \bibinfo {author} {\bibfnamefont
  {J.}~\bibnamefont {Hone}}, \bibinfo {author} {\bibfnamefont {P.}~\bibnamefont
  {Kim}}, \bibinfo {author} {\bibfnamefont {H.~L.}\ \bibnamefont {Stormer}},
  \bibinfo {author} {\bibfnamefont {V.}~\bibnamefont {Yakimenko}}, \ and\
  \bibinfo {author} {\bibfnamefont {G.}~\bibnamefont {Flynn}},\ }\href
  {\doibase 10.1021/nl803087x} {\bibfield  {journal} {\bibinfo  {journal} {Nano
  Lett.}\ }\textbf {\bibinfo {volume} {9}},\ \bibinfo {pages} {332} (\bibinfo
  {year} {2009})}\BibitemShut {NoStop}%
\bibitem [{\citenamefont {Georgiou}\ \emph {et~al.}(2011)\citenamefont
  {Georgiou}, \citenamefont {Britnell}, \citenamefont {Blake}, \citenamefont
  {Gorbachev}, \citenamefont {Gholinia}, \citenamefont {Geim}, \citenamefont
  {Casiraghi},\ and\ \citenamefont {Novoselov}}]{Georgiou2011}%
  \BibitemOpen
  \bibfield  {author} {\bibinfo {author} {\bibfnamefont {T.}~\bibnamefont
  {Georgiou}}, \bibinfo {author} {\bibfnamefont {L.}~\bibnamefont {Britnell}},
  \bibinfo {author} {\bibfnamefont {P.}~\bibnamefont {Blake}}, \bibinfo
  {author} {\bibfnamefont {R.~V.}\ \bibnamefont {Gorbachev}}, \bibinfo {author}
  {\bibfnamefont {A.}~\bibnamefont {Gholinia}}, \bibinfo {author}
  {\bibfnamefont {A.~K.}\ \bibnamefont {Geim}}, \bibinfo {author}
  {\bibfnamefont {C.}~\bibnamefont {Casiraghi}}, \ and\ \bibinfo {author}
  {\bibfnamefont {K.~S.}\ \bibnamefont {Novoselov}},\ }\href {\doibase
  10.1063/1.3631632} {\bibfield  {journal} {\bibinfo  {journal} {Appl. Phys.
  Lett.}\ }\textbf {\bibinfo {volume} {99}},\ \bibinfo {pages} {093103}
  (\bibinfo {year} {2011})}\BibitemShut {NoStop}%
\bibitem [{\citenamefont {Lim}\ \emph {et~al.}(2013)\citenamefont {Lim},
  \citenamefont {Sorkin}, \citenamefont {Bao}, \citenamefont {Li},
  \citenamefont {Zhang}, \citenamefont {Nesladek},\ and\ \citenamefont
  {Loh}}]{Lim2013}%
  \BibitemOpen
  \bibfield  {author} {\bibinfo {author} {\bibfnamefont {C.~H. Y.~X.}\
  \bibnamefont {Lim}}, \bibinfo {author} {\bibfnamefont {A.}~\bibnamefont
  {Sorkin}}, \bibinfo {author} {\bibfnamefont {Q.}~\bibnamefont {Bao}},
  \bibinfo {author} {\bibfnamefont {A.}~\bibnamefont {Li}}, \bibinfo {author}
  {\bibfnamefont {K.}~\bibnamefont {Zhang}}, \bibinfo {author} {\bibfnamefont
  {M.}~\bibnamefont {Nesladek}}, \ and\ \bibinfo {author} {\bibfnamefont
  {K.~P.}\ \bibnamefont {Loh}},\ }\href {\doibase 10.1038/ncomms2579}
  {\bibfield  {journal} {\bibinfo  {journal} {Nat. Commun.}\ }\textbf {\bibinfo
  {volume} {4}},\ \bibinfo {pages} {1556} (\bibinfo {year} {2013})}\BibitemShut
  {NoStop}%
\bibitem [{\citenamefont {Pan}\ \emph {et~al.}(2012)\citenamefont {Pan},
  \citenamefont {Xiao}, \citenamefont {Zhu}, \citenamefont {Yu}, \citenamefont
  {Zhang}, \citenamefont {Ni}, \citenamefont {Watanabe}, \citenamefont
  {Taniguchi}, \citenamefont {Shi},\ and\ \citenamefont {Wang}}]{Pan2012}%
  \BibitemOpen
  \bibfield  {author} {\bibinfo {author} {\bibfnamefont {W.}~\bibnamefont
  {Pan}}, \bibinfo {author} {\bibfnamefont {J.}~\bibnamefont {Xiao}}, \bibinfo
  {author} {\bibfnamefont {J.}~\bibnamefont {Zhu}}, \bibinfo {author}
  {\bibfnamefont {C.}~\bibnamefont {Yu}}, \bibinfo {author} {\bibfnamefont
  {G.}~\bibnamefont {Zhang}}, \bibinfo {author} {\bibfnamefont
  {Z.}~\bibnamefont {Ni}}, \bibinfo {author} {\bibfnamefont {K.}~\bibnamefont
  {Watanabe}}, \bibinfo {author} {\bibfnamefont {T.}~\bibnamefont {Taniguchi}},
  \bibinfo {author} {\bibfnamefont {Y.}~\bibnamefont {Shi}}, \ and\ \bibinfo
  {author} {\bibfnamefont {X.}~\bibnamefont {Wang}},\ }\href {\doibase
  10.1038/srep00893} {\bibfield  {journal} {\bibinfo  {journal} {Sci. Rep.}\
  }\textbf {\bibinfo {volume} {2}},\ \bibinfo {pages} {893} (\bibinfo {year}
  {2012})}\BibitemShut {NoStop}%
\bibitem [{\citenamefont {Yuk}\ \emph {et~al.}(2012)\citenamefont {Yuk},
  \citenamefont {Park}, \citenamefont {Ercius}, \citenamefont {Kim},
  \citenamefont {Hellebusch}, \citenamefont {Crommie}, \citenamefont {Lee},
  \citenamefont {Zettl},\ and\ \citenamefont {Alivisatos}}]{Yuk2012}%
  \BibitemOpen
  \bibfield  {author} {\bibinfo {author} {\bibfnamefont {J.~M.}\ \bibnamefont
  {Yuk}}, \bibinfo {author} {\bibfnamefont {J.}~\bibnamefont {Park}}, \bibinfo
  {author} {\bibfnamefont {P.}~\bibnamefont {Ercius}}, \bibinfo {author}
  {\bibfnamefont {K.}~\bibnamefont {Kim}}, \bibinfo {author} {\bibfnamefont
  {D.~J.}\ \bibnamefont {Hellebusch}}, \bibinfo {author} {\bibfnamefont
  {M.~F.}\ \bibnamefont {Crommie}}, \bibinfo {author} {\bibfnamefont {J.~Y.}\
  \bibnamefont {Lee}}, \bibinfo {author} {\bibfnamefont {A.}~\bibnamefont
  {Zettl}}, \ and\ \bibinfo {author} {\bibfnamefont {A.~P.}\ \bibnamefont
  {Alivisatos}},\ }\href {\doibase 10.1126/science.1217654} {\bibfield
  {journal} {\bibinfo  {journal} {Science}\ }\textbf {\bibinfo {volume}
  {336}},\ \bibinfo {pages} {61} (\bibinfo {year} {2012})}\BibitemShut
  {NoStop}%
\bibitem [{\citenamefont {Guinea}\ \emph {et~al.}(2009)\citenamefont {Guinea},
  \citenamefont {Katsnelson},\ and\ \citenamefont {Geim}}]{Guinea2009}%
  \BibitemOpen
  \bibfield  {author} {\bibinfo {author} {\bibfnamefont {F.}~\bibnamefont
  {Guinea}}, \bibinfo {author} {\bibfnamefont {M.~I.}\ \bibnamefont
  {Katsnelson}}, \ and\ \bibinfo {author} {\bibfnamefont {A.~K.}\ \bibnamefont
  {Geim}},\ }\href {\doibase 10.1038/nphys1420} {\bibfield  {journal} {\bibinfo
   {journal} {Nat. Phys.}\ }\textbf {\bibinfo {volume} {6}},\ \bibinfo {pages}
  {30} (\bibinfo {year} {2009})}\BibitemShut {NoStop}%
\bibitem [{\citenamefont {Levy}\ \emph {et~al.}(2010)\citenamefont {Levy},
  \citenamefont {Burke}, \citenamefont {Meaker}, \citenamefont {Panlasigui},
  \citenamefont {Zettl}, \citenamefont {Guinea}, \citenamefont {{Castro
  Neto}},\ and\ \citenamefont {Crommie}}]{Levy2010}%
  \BibitemOpen
  \bibfield  {author} {\bibinfo {author} {\bibfnamefont {N.}~\bibnamefont
  {Levy}}, \bibinfo {author} {\bibfnamefont {S.~A.}\ \bibnamefont {Burke}},
  \bibinfo {author} {\bibfnamefont {K.~L.}\ \bibnamefont {Meaker}}, \bibinfo
  {author} {\bibfnamefont {M.}~\bibnamefont {Panlasigui}}, \bibinfo {author}
  {\bibfnamefont {A.}~\bibnamefont {Zettl}}, \bibinfo {author} {\bibfnamefont
  {F.}~\bibnamefont {Guinea}}, \bibinfo {author} {\bibfnamefont {A.~H.}\
  \bibnamefont {{Castro Neto}}}, \ and\ \bibinfo {author} {\bibfnamefont
  {M.~F.}\ \bibnamefont {Crommie}},\ }\href {\doibase 10.1126/science.1191700}
  {\bibfield  {journal} {\bibinfo  {journal} {Science}\ }\textbf {\bibinfo
  {volume} {329}},\ \bibinfo {pages} {544} (\bibinfo {year}
  {2010})}\BibitemShut {NoStop}%
\bibitem [{\citenamefont {Lu}\ \emph {et~al.}(2012)\citenamefont {Lu},
  \citenamefont {Castro~Neto},\ and\ \citenamefont {Loh}}]{Lu2012}%
  \BibitemOpen
  \bibfield  {author} {\bibinfo {author} {\bibfnamefont {J.}~\bibnamefont
  {Lu}}, \bibinfo {author} {\bibfnamefont {A.~H.}\ \bibnamefont {Castro~Neto}},
  \ and\ \bibinfo {author} {\bibfnamefont {K.~P.}\ \bibnamefont {Loh}},\ }\href
  {\doibase 10.1038/ncomms1818} {\bibfield  {journal} {\bibinfo  {journal}
  {Nat. Commun.}\ }\textbf {\bibinfo {volume} {3}},\ \bibinfo {pages} {823}
  (\bibinfo {year} {2012})}\BibitemShut {NoStop}%
\bibitem [{\citenamefont {Scherzer}(1983)}]{Scherzer1983}%
  \BibitemOpen
  \bibfield  {author} {\bibinfo {author} {\bibfnamefont {B.~M.~U.}\
  \bibnamefont {Scherzer}},\ }\href@noop {} {\emph {\bibinfo {title} {Top.
  Appl. Phys.}}},\ edited by\ \bibinfo {editor} {\bibfnamefont
  {R.}~\bibnamefont {Behrisch}},\ Vol.~\bibinfo {volume} {52}\ (\bibinfo
  {publisher} {Springer Berlin},\ \bibinfo {year} {1983})\ p.\ \bibinfo {pages}
  {271}\BibitemShut {NoStop}%
\bibitem [{\citenamefont {Bruel}(1995)}]{Bruel1995}%
  \BibitemOpen
  \bibfield  {author} {\bibinfo {author} {\bibfnamefont {M.}~\bibnamefont
  {Bruel}},\ }\href {\doibase 10.1049/el:19950805} {\bibfield  {journal}
  {\bibinfo  {journal} {Electron. Lett.}\ }\textbf {\bibinfo {volume} {31}},\
  \bibinfo {pages} {1201} (\bibinfo {year} {1995})}\BibitemShut {NoStop}%
\bibitem [{\citenamefont {Alimov}\ \emph {et~al.}(1995)\citenamefont {Alimov},
  \citenamefont {Scherzer}, \citenamefont {Chernikov},\ and\ \citenamefont
  {Ullmaier}}]{Alimov1995}%
  \BibitemOpen
  \bibfield  {author} {\bibinfo {author} {\bibfnamefont {V.~K.}\ \bibnamefont
  {Alimov}}, \bibinfo {author} {\bibfnamefont {B.~M.~U.}\ \bibnamefont
  {Scherzer}}, \bibinfo {author} {\bibfnamefont {V.~N.}\ \bibnamefont
  {Chernikov}}, \ and\ \bibinfo {author} {\bibfnamefont {H.}~\bibnamefont
  {Ullmaier}},\ }\href {\doibase 10.1063/1.360664} {\bibfield  {journal}
  {\bibinfo  {journal} {J. Appl. Phys.}\ }\textbf {\bibinfo {volume} {78}},\
  \bibinfo {pages} {137} (\bibinfo {year} {1995})}\BibitemShut {NoStop}%
\bibitem [{\citenamefont {Chernikov}\ \emph {et~al.}(1996)\citenamefont
  {Chernikov}, \citenamefont {Kesternich},\ and\ \citenamefont
  {Ullmaier}}]{Chernikov1996}%
  \BibitemOpen
  \bibfield  {author} {\bibinfo {author} {\bibfnamefont {V.~N.}\ \bibnamefont
  {Chernikov}}, \bibinfo {author} {\bibfnamefont {W.}~\bibnamefont
  {Kesternich}}, \ and\ \bibinfo {author} {\bibfnamefont {H.}~\bibnamefont
  {Ullmaier}},\ }\href {\doibase 10.1016/0022-3115(95)00157-3} {\bibfield
  {journal} {\bibinfo  {journal} {J. Nucl. Mater.}\ }\textbf {\bibinfo {volume}
  {227}},\ \bibinfo {pages} {157} (\bibinfo {year} {1996})}\BibitemShut
  {NoStop}%
\bibitem [{\citenamefont {Bangert}\ \emph {et~al.}(2013)\citenamefont
  {Bangert}, \citenamefont {Pierce}, \citenamefont {Kepaptsoglou},
  \citenamefont {Ramasse}, \citenamefont {Zan}, \citenamefont {Gass},
  \citenamefont {Van~den Berg}, \citenamefont {Boothroyd}, \citenamefont
  {Amani},\ and\ \citenamefont {Hofs\"{a}ss}}]{Bangert2013}%
  \BibitemOpen
  \bibfield  {author} {\bibinfo {author} {\bibfnamefont {U.}~\bibnamefont
  {Bangert}}, \bibinfo {author} {\bibfnamefont {W.}~\bibnamefont {Pierce}},
  \bibinfo {author} {\bibfnamefont {D.~M.}\ \bibnamefont {Kepaptsoglou}},
  \bibinfo {author} {\bibfnamefont {Q.}~\bibnamefont {Ramasse}}, \bibinfo
  {author} {\bibfnamefont {R.}~\bibnamefont {Zan}}, \bibinfo {author}
  {\bibfnamefont {M.~H.}\ \bibnamefont {Gass}}, \bibinfo {author}
  {\bibfnamefont {J.~A.}\ \bibnamefont {Van~den Berg}}, \bibinfo {author}
  {\bibfnamefont {C.~B.}\ \bibnamefont {Boothroyd}}, \bibinfo {author}
  {\bibfnamefont {J.}~\bibnamefont {Amani}}, \ and\ \bibinfo {author}
  {\bibfnamefont {H.}~\bibnamefont {Hofs\"{a}ss}},\ }\href@noop {} {\bibfield
  {journal} {\bibinfo  {journal} {Nano Lett.}\ }\textbf {\bibinfo {volume}
  {13}},\ \bibinfo {pages} {4902} (\bibinfo {year} {2013})}\BibitemShut
  {NoStop}%
\bibitem [{\citenamefont {Telychko}\ \emph {et~al.}(2014)\citenamefont
  {Telychko}, \citenamefont {Mutombo}, \citenamefont {Ondr\'{a}\v{c}ek},
  \citenamefont {Hapala}, \citenamefont {Bocquet}, \citenamefont
  {Koloren\v{c}}, \citenamefont {Vondr\'{a}\v{c}ek}, \citenamefont
  {Jel\'{i}nek},\ and\ \citenamefont {\v{S}vec}}]{Telychko2014}%
  \BibitemOpen
  \bibfield  {author} {\bibinfo {author} {\bibfnamefont {M.}~\bibnamefont
  {Telychko}}, \bibinfo {author} {\bibfnamefont {P.}~\bibnamefont {Mutombo}},
  \bibinfo {author} {\bibfnamefont {M.}~\bibnamefont {Ondr\'{a}\v{c}ek}},
  \bibinfo {author} {\bibfnamefont {P.}~\bibnamefont {Hapala}}, \bibinfo
  {author} {\bibfnamefont {F.~C.}\ \bibnamefont {Bocquet}}, \bibinfo {author}
  {\bibfnamefont {J.}~\bibnamefont {Koloren\v{c}}}, \bibinfo {author}
  {\bibfnamefont {M.}~\bibnamefont {Vondr\'{a}\v{c}ek}}, \bibinfo {author}
  {\bibfnamefont {P.}~\bibnamefont {Jel\'{i}nek}}, \ and\ \bibinfo {author}
  {\bibfnamefont {M.}~\bibnamefont {\v{S}vec}},\ }\href@noop {} {\bibfield
  {journal} {\bibinfo  {journal} {ACS Nano}\ }\textbf {\bibinfo {volume} {8}},\
  \bibinfo {pages} {7318} (\bibinfo {year} {2014})}\BibitemShut {NoStop}%
\bibitem [{\citenamefont {Cun}\ \emph {et~al.}(2013)\citenamefont {Cun},
  \citenamefont {Iannuzzi}, \citenamefont {Hemmi}, \citenamefont {Roth},
  \citenamefont {Osterwalder},\ and\ \citenamefont {Greber}}]{Cun2013}%
  \BibitemOpen
  \bibfield  {author} {\bibinfo {author} {\bibfnamefont {H.}~\bibnamefont
  {Cun}}, \bibinfo {author} {\bibfnamefont {M.}~\bibnamefont {Iannuzzi}},
  \bibinfo {author} {\bibfnamefont {A.}~\bibnamefont {Hemmi}}, \bibinfo
  {author} {\bibfnamefont {S.}~\bibnamefont {Roth}}, \bibinfo {author}
  {\bibfnamefont {J.}~\bibnamefont {Osterwalder}}, \ and\ \bibinfo {author}
  {\bibfnamefont {T.}~\bibnamefont {Greber}},\ }\href {\doibase
  10.1021/nl400449y} {\bibfield  {journal} {\bibinfo  {journal} {Nano Lett.}\
  }\textbf {\bibinfo {volume} {13}},\ \bibinfo {pages} {2098} (\bibinfo {year}
  {2013})}\BibitemShut {NoStop}%
\bibitem [{\citenamefont {Cun}\ \emph {et~al.}(2014{\natexlab{a}})\citenamefont
  {Cun}, \citenamefont {Iannuzzi}, \citenamefont {Hemmi}, \citenamefont
  {Osterwalder},\ and\ \citenamefont {Greber}}]{Cun2014a}%
  \BibitemOpen
  \bibfield  {author} {\bibinfo {author} {\bibfnamefont {H.}~\bibnamefont
  {Cun}}, \bibinfo {author} {\bibfnamefont {M.}~\bibnamefont {Iannuzzi}},
  \bibinfo {author} {\bibfnamefont {A.}~\bibnamefont {Hemmi}}, \bibinfo
  {author} {\bibfnamefont {J.}~\bibnamefont {Osterwalder}}, \ and\ \bibinfo
  {author} {\bibfnamefont {T.}~\bibnamefont {Greber}},\ }\href {\doibase
  10.1021/nn405907a} {\bibfield  {journal} {\bibinfo  {journal} {ACS Nano}\
  }\textbf {\bibinfo {volume} {8}},\ \bibinfo {pages} {1014} (\bibinfo {year}
  {2014}{\natexlab{a}})}\BibitemShut {NoStop}%
\bibitem [{\citenamefont {Cun}\ \emph {et~al.}(2014{\natexlab{b}})\citenamefont
  {Cun}, \citenamefont {Iannuzzi}, \citenamefont {Hemmi}, \citenamefont
  {Osterwalder},\ and\ \citenamefont {Greber}}]{Cun2014b}%
  \BibitemOpen
  \bibfield  {author} {\bibinfo {author} {\bibfnamefont {H.}~\bibnamefont
  {Cun}}, \bibinfo {author} {\bibfnamefont {M.}~\bibnamefont {Iannuzzi}},
  \bibinfo {author} {\bibfnamefont {A.}~\bibnamefont {Hemmi}}, \bibinfo
  {author} {\bibfnamefont {J.}~\bibnamefont {Osterwalder}}, \ and\ \bibinfo
  {author} {\bibfnamefont {T.}~\bibnamefont {Greber}},\ }\href {\doibase
  10.1021/nn502645w} {\bibfield  {journal} {\bibinfo  {journal} {ACS Nano}\
  }\textbf {\bibinfo {volume} {8}},\ \bibinfo {pages} {7423} (\bibinfo {year}
  {2014}{\natexlab{b}})}\BibitemShut {NoStop}%
\bibitem [{\citenamefont {Cun}\ \emph {et~al.}(2015)\citenamefont {Cun},
  \citenamefont {Iannuzzi}, \citenamefont {Hemmi}, \citenamefont
  {Osterwalder},\ and\ \citenamefont {Greber}}]{Cun2015}%
  \BibitemOpen
  \bibfield  {author} {\bibinfo {author} {\bibfnamefont {H.}~\bibnamefont
  {Cun}}, \bibinfo {author} {\bibfnamefont {M.}~\bibnamefont {Iannuzzi}},
  \bibinfo {author} {\bibfnamefont {A.}~\bibnamefont {Hemmi}}, \bibinfo
  {author} {\bibfnamefont {J.}~\bibnamefont {Osterwalder}}, \ and\ \bibinfo
  {author} {\bibfnamefont {T.}~\bibnamefont {Greber}},\ }\href {\doibase
  10.1016/j.susc.2014.11.004} {\bibfield  {journal} {\bibinfo  {journal} {Surf.
  Sci.}\ }\textbf {\bibinfo {volume} {634}},\ \bibinfo {pages} {95} (\bibinfo
  {year} {2015})}\BibitemShut {NoStop}%
\bibitem [{\citenamefont {Herbig}\ \emph {et~al.}(2015)\citenamefont {Herbig},
  \citenamefont {{\AA}hlgren}, \citenamefont {Schr\"{o}der}, \citenamefont
  {Mart\'{\i}nez-Galera}, \citenamefont {Arman}, \citenamefont {Jolie},
  \citenamefont {Busse}, \citenamefont {Kotakoski}, \citenamefont {Knudsen},
  \citenamefont {Krasheninnikov},\ and\ \citenamefont {Michely}}]{Herbig2015}%
  \BibitemOpen
  \bibfield  {author} {\bibinfo {author} {\bibfnamefont {C.}~\bibnamefont
  {Herbig}}, \bibinfo {author} {\bibfnamefont {E.~H.}\ \bibnamefont
  {{\AA}hlgren}}, \bibinfo {author} {\bibfnamefont {U.~A.}\ \bibnamefont
  {Schr\"{o}der}}, \bibinfo {author} {\bibfnamefont {A.~J.}\ \bibnamefont
  {Mart\'{\i}nez-Galera}}, \bibinfo {author} {\bibfnamefont {M.~A.}\
  \bibnamefont {Arman}}, \bibinfo {author} {\bibfnamefont {W.}~\bibnamefont
  {Jolie}}, \bibinfo {author} {\bibfnamefont {C.}~\bibnamefont {Busse}},
  \bibinfo {author} {\bibfnamefont {J.}~\bibnamefont {Kotakoski}}, \bibinfo
  {author} {\bibfnamefont {J.}~\bibnamefont {Knudsen}}, \bibinfo {author}
  {\bibfnamefont {A.~V.}\ \bibnamefont {Krasheninnikov}}, \ and\ \bibinfo
  {author} {\bibfnamefont {T.}~\bibnamefont {Michely}},\ }\href@noop {}
  {\bibfield  {journal} {\bibinfo  {journal} {ACS Nano}\ }\textbf {\bibinfo
  {volume} {9}},\ \bibinfo {pages} {4664} (\bibinfo {year} {2015})}\BibitemShut
  {NoStop}%
\bibitem [{\citenamefont {Herbig}\ \emph {et~al.}(2014)\citenamefont {Herbig},
  \citenamefont {{\AA}hlgren}, \citenamefont {Jolie}, \citenamefont {Busse},
  \citenamefont {Kotakoski}, \citenamefont {Krasheninnikov},\ and\
  \citenamefont {Michely}}]{Herbig2014}%
  \BibitemOpen
  \bibfield  {author} {\bibinfo {author} {\bibfnamefont {C.}~\bibnamefont
  {Herbig}}, \bibinfo {author} {\bibfnamefont {E.~H.}\ \bibnamefont
  {{\AA}hlgren}}, \bibinfo {author} {\bibfnamefont {W.}~\bibnamefont {Jolie}},
  \bibinfo {author} {\bibfnamefont {C.}~\bibnamefont {Busse}}, \bibinfo
  {author} {\bibfnamefont {J.}~\bibnamefont {Kotakoski}}, \bibinfo {author}
  {\bibfnamefont {A.~V.}\ \bibnamefont {Krasheninnikov}}, \ and\ \bibinfo
  {author} {\bibfnamefont {T.}~\bibnamefont {Michely}},\ }\href {\doibase
  10.1021/nn503874n} {\bibfield  {journal} {\bibinfo  {journal} {ACS Nano}\
  }\textbf {\bibinfo {volume} {8}},\ \bibinfo {pages} {12208} (\bibinfo {year}
  {2014})}\BibitemShut {NoStop}%
\bibitem [{\citenamefont {van Gastel}\ \emph {et~al.}(2009)\citenamefont {van
  Gastel}, \citenamefont {N'Diaye}, \citenamefont {Wall}, \citenamefont
  {Coraux}, \citenamefont {Busse}, \citenamefont {Buckanie}, \citenamefont
  {{Meyer zu Heringdorf}}, \citenamefont {{Horn von Hoegen}}, \citenamefont
  {Michely},\ and\ \citenamefont {Poelsema}}]{vanGastel2009}%
  \BibitemOpen
  \bibfield  {author} {\bibinfo {author} {\bibfnamefont {R.}~\bibnamefont {van
  Gastel}}, \bibinfo {author} {\bibfnamefont {A.~T.}\ \bibnamefont {N'Diaye}},
  \bibinfo {author} {\bibfnamefont {D.}~\bibnamefont {Wall}}, \bibinfo {author}
  {\bibfnamefont {J.}~\bibnamefont {Coraux}}, \bibinfo {author} {\bibfnamefont
  {C.}~\bibnamefont {Busse}}, \bibinfo {author} {\bibfnamefont {N.~M.}\
  \bibnamefont {Buckanie}}, \bibinfo {author} {\bibfnamefont {F.-J.}\
  \bibnamefont {{Meyer zu Heringdorf}}}, \bibinfo {author} {\bibfnamefont
  {M.}~\bibnamefont {{Horn von Hoegen}}}, \bibinfo {author} {\bibfnamefont
  {T.}~\bibnamefont {Michely}}, \ and\ \bibinfo {author} {\bibfnamefont
  {B.}~\bibnamefont {Poelsema}},\ }\href {\doibase 10.1063/1.3225554}
  {\bibfield  {journal} {\bibinfo  {journal} {Appl. Phys. Lett.}\ }\textbf
  {\bibinfo {volume} {95}},\ \bibinfo {pages} {121901} (\bibinfo {year}
  {2009})}\BibitemShut {NoStop}%
\bibitem [{Sup()}]{SuppInfo}%
  \BibitemOpen
  \href@noop {} {}\bibinfo {note} {See Supplemental Material at [URL will be
  inserted by publisher] for plots of the Xe depth distribution obtained by MD
  simulations (Fig. S1), supercell geometries used for DFT calculation and
  results of DFT calculations for Xe atom diffusion through the edge of a Gr
  vacancy (Figs. S2 and S3), additional XP-spectra after 3\,keV Xe${^+}$ ion
  irradiation of bare Ir(111) and Gr/Ir(111) (Fig. S4), STM topographs of the
  latter (Fig. S5), and a sample morphology after sputter cleaning together
  with insufficient annealing prior to Gr growth (Fig. S6).}\BibitemShut
  {Stop}%
\bibitem [{\citenamefont {Nordlund}\ \emph {et~al.}(1998)\citenamefont
  {Nordlund}, \citenamefont {Ghaly}, \citenamefont {Averback}, \citenamefont
  {Caturla}, \citenamefont {{Diaz de la Rubia}},\ and\ \citenamefont
  {Tarus}}]{Nordlund1998}%
  \BibitemOpen
  \bibfield  {author} {\bibinfo {author} {\bibfnamefont {K.}~\bibnamefont
  {Nordlund}}, \bibinfo {author} {\bibfnamefont {M.}~\bibnamefont {Ghaly}},
  \bibinfo {author} {\bibfnamefont {R.~S.}\ \bibnamefont {Averback}}, \bibinfo
  {author} {\bibfnamefont {M.}~\bibnamefont {Caturla}}, \bibinfo {author}
  {\bibfnamefont {T.}~\bibnamefont {{Diaz de la Rubia}}}, \ and\ \bibinfo
  {author} {\bibfnamefont {J.}~\bibnamefont {Tarus}},\ }\href {\doibase
  10.1103/PhysRevB.57.7556} {\bibfield  {journal} {\bibinfo  {journal} {Phys.
  Rev. B}\ }\textbf {\bibinfo {volume} {57}},\ \bibinfo {pages} {7556}
  (\bibinfo {year} {1998})}\BibitemShut {NoStop}%
\bibitem [{\citenamefont {Albe}\ \emph {et~al.}(2002)\citenamefont {Albe},
  \citenamefont {Nordlund},\ and\ \citenamefont {Averback}}]{Albe2002}%
  \BibitemOpen
  \bibfield  {author} {\bibinfo {author} {\bibfnamefont {K.}~\bibnamefont
  {Albe}}, \bibinfo {author} {\bibfnamefont {K.}~\bibnamefont {Nordlund}}, \
  and\ \bibinfo {author} {\bibfnamefont {R.~S.}\ \bibnamefont {Averback}},\
  }\href {\doibase 10.1103/PhysRevB.65.195124} {\bibfield  {journal} {\bibinfo
  {journal} {Phys. Rev. B}\ }\textbf {\bibinfo {volume} {65}},\ \bibinfo
  {pages} {195124} (\bibinfo {year} {2002})}\BibitemShut {NoStop}%
\bibitem [{\citenamefont {Berendsen}\ \emph {et~al.}(1984)\citenamefont
  {Berendsen}, \citenamefont {Postma}, \citenamefont {van Gunsteren},
  \citenamefont {DiNola},\ and\ \citenamefont {Haak}}]{Berendsen1984}%
  \BibitemOpen
  \bibfield  {author} {\bibinfo {author} {\bibfnamefont {H.~J.~C.}\
  \bibnamefont {Berendsen}}, \bibinfo {author} {\bibfnamefont {J.~P.~M.}\
  \bibnamefont {Postma}}, \bibinfo {author} {\bibfnamefont {W.~F.}\
  \bibnamefont {van Gunsteren}}, \bibinfo {author} {\bibfnamefont
  {A.}~\bibnamefont {DiNola}}, \ and\ \bibinfo {author} {\bibfnamefont {J.~R.}\
  \bibnamefont {Haak}},\ }\href {\doibase 10.1063/1.448118} {\bibfield
  {journal} {\bibinfo  {journal} {J. Chem. Phys.}\ }\textbf {\bibinfo {volume}
  {81}},\ \bibinfo {pages} {3684} (\bibinfo {year} {1984})}\BibitemShut
  {NoStop}%
\bibitem [{\citenamefont {Sutter}\ \emph {et~al.}(2009)\citenamefont {Sutter},
  \citenamefont {Sadowski},\ and\ \citenamefont {Sutter}}]{Sutter2009}%
  \BibitemOpen
  \bibfield  {author} {\bibinfo {author} {\bibfnamefont {P.}~\bibnamefont
  {Sutter}}, \bibinfo {author} {\bibfnamefont {J.~T.}\ \bibnamefont
  {Sadowski}}, \ and\ \bibinfo {author} {\bibfnamefont {E.}~\bibnamefont
  {Sutter}},\ }\href@noop {} {\bibfield  {journal} {\bibinfo  {journal} {Phys.
  Rev. B}\ }\textbf {\bibinfo {volume} {80}},\ \bibinfo {pages} {245411}
  (\bibinfo {year} {2009})}\BibitemShut {NoStop}%
\bibitem [{\citenamefont {Brenner}\ \emph {et~al.}(2002)\citenamefont
  {Brenner}, \citenamefont {Shenderova}, \citenamefont {Harrison},
  \citenamefont {Stuart}, \citenamefont {Ni},\ and\ \citenamefont
  {Sinnott}}]{Brenner2002}%
  \BibitemOpen
  \bibfield  {author} {\bibinfo {author} {\bibfnamefont {D.~W.}\ \bibnamefont
  {Brenner}}, \bibinfo {author} {\bibfnamefont {O.~A.}\ \bibnamefont
  {Shenderova}}, \bibinfo {author} {\bibfnamefont {J.~A.}\ \bibnamefont
  {Harrison}}, \bibinfo {author} {\bibfnamefont {S.~J.}\ \bibnamefont
  {Stuart}}, \bibinfo {author} {\bibfnamefont {B.}~\bibnamefont {Ni}}, \ and\
  \bibinfo {author} {\bibfnamefont {S.~B.}\ \bibnamefont {Sinnott}},\ }\href
  {\doibase 10.1088/0953-8984/14/4/312} {\bibfield  {journal} {\bibinfo
  {journal} {J. Phys. Condens. Matter}\ }\textbf {\bibinfo {volume} {14}},\
  \bibinfo {pages} {783} (\bibinfo {year} {2002})}\BibitemShut {NoStop}%
\bibitem [{\citenamefont {Ziegler}\ \emph {et~al.}(1985)\citenamefont
  {Ziegler}, \citenamefont {Biersack},\ and\ \citenamefont {Littmark}}]{ZBL}%
  \BibitemOpen
  \bibfield  {author} {\bibinfo {author} {\bibfnamefont {J.~F.}\ \bibnamefont
  {Ziegler}}, \bibinfo {author} {\bibfnamefont {J.~P.}\ \bibnamefont
  {Biersack}}, \ and\ \bibinfo {author} {\bibfnamefont {U.}~\bibnamefont
  {Littmark}},\ }in\ \href {\doibase 10.1007/978-1-4615-8103-1\_3} {\emph
  {\bibinfo {booktitle} {Treatise Heavy-Ion Sci.}}},\ \bibinfo {editor} {edited
  by\ \bibinfo {editor} {\bibfnamefont {D.}~\bibnamefont {Bromley}}}\ (\bibinfo
   {publisher} {Springer US},\ \bibinfo {address} {New York},\ \bibinfo {year}
  {1985})\ pp.\ \bibinfo {pages} {93--129}\BibitemShut {NoStop}%
\bibitem [{\citenamefont {Kresse}\ and\ \citenamefont
  {Furthm\"{u}ller}(1996{\natexlab{a}})}]{VASP1}%
  \BibitemOpen
  \bibfield  {author} {\bibinfo {author} {\bibfnamefont {G.}~\bibnamefont
  {Kresse}}\ and\ \bibinfo {author} {\bibfnamefont {J.}~\bibnamefont
  {Furthm\"{u}ller}},\ }\href {\doibase 10.1016/0927-0256(96)00008-0}
  {\bibfield  {journal} {\bibinfo  {journal} {Comput. Mater. Sci.}\ }\textbf
  {\bibinfo {volume} {6}},\ \bibinfo {pages} {15} (\bibinfo {year}
  {1996}{\natexlab{a}})}\BibitemShut {NoStop}%
\bibitem [{\citenamefont {Kresse}\ and\ \citenamefont
  {Furthm\"{u}ller}(1996{\natexlab{b}})}]{VASP2}%
  \BibitemOpen
  \bibfield  {author} {\bibinfo {author} {\bibfnamefont {G.}~\bibnamefont
  {Kresse}}\ and\ \bibinfo {author} {\bibfnamefont {J.}~\bibnamefont
  {Furthm\"{u}ller}},\ }\href {\doibase 10.1103/PhysRevB.54.11169} {\bibfield
  {journal} {\bibinfo  {journal} {Phys. Rev. B}\ }\textbf {\bibinfo {volume}
  {54}},\ \bibinfo {pages} {11169} (\bibinfo {year}
  {1996}{\natexlab{b}})}\BibitemShut {NoStop}%
\bibitem [{\citenamefont {Bl\"{o}chl}(1994)}]{PAW2}%
  \BibitemOpen
  \bibfield  {author} {\bibinfo {author} {\bibfnamefont {P.~E.}\ \bibnamefont
  {Bl\"{o}chl}},\ }\href {\doibase 10.1103/PhysRevB.50.17953} {\bibfield
  {journal} {\bibinfo  {journal} {Phys. Rev. B}\ }\textbf {\bibinfo {volume}
  {50}},\ \bibinfo {pages} {17953} (\bibinfo {year} {1994})}\BibitemShut
  {NoStop}%
\bibitem [{\citenamefont {Bj\"{o}rkman}(2012)}]{Bjorkman2012}%
  \BibitemOpen
  \bibfield  {author} {\bibinfo {author} {\bibfnamefont {T.}~\bibnamefont
  {Bj\"{o}rkman}},\ }\href {\doibase 10.1103/PhysRevB.86.165109} {\bibfield
  {journal} {\bibinfo  {journal} {Phys. Rev. B}\ }\textbf {\bibinfo {volume}
  {86}},\ \bibinfo {pages} {165109} (\bibinfo {year} {2012})}\BibitemShut
  {NoStop}%
\bibitem [{\citenamefont {Standop}\ \emph {et~al.}(2013)\citenamefont
  {Standop}, \citenamefont {Lehtinen}, \citenamefont {Herbig}, \citenamefont
  {Lewes-Malandrakis}, \citenamefont {Craes}, \citenamefont {Kotakoski},
  \citenamefont {Michely}, \citenamefont {Krasheninnikov},\ and\ \citenamefont
  {Busse}}]{Standop2013}%
  \BibitemOpen
  \bibfield  {author} {\bibinfo {author} {\bibfnamefont {S.}~\bibnamefont
  {Standop}}, \bibinfo {author} {\bibfnamefont {O.}~\bibnamefont {Lehtinen}},
  \bibinfo {author} {\bibfnamefont {C.}~\bibnamefont {Herbig}}, \bibinfo
  {author} {\bibfnamefont {G.}~\bibnamefont {Lewes-Malandrakis}}, \bibinfo
  {author} {\bibfnamefont {F.}~\bibnamefont {Craes}}, \bibinfo {author}
  {\bibfnamefont {J.}~\bibnamefont {Kotakoski}}, \bibinfo {author}
  {\bibfnamefont {T.}~\bibnamefont {Michely}}, \bibinfo {author} {\bibfnamefont
  {A.~V.}\ \bibnamefont {Krasheninnikov}}, \ and\ \bibinfo {author}
  {\bibfnamefont {C.}~\bibnamefont {Busse}},\ }\href {\doibase
  10.1021/nl304659n} {\bibfield  {journal} {\bibinfo  {journal} {Nano Lett.}\
  }\textbf {\bibinfo {volume} {13}},\ \bibinfo {pages} {1948} (\bibinfo {year}
  {2013})}\BibitemShut {NoStop}%
\bibitem [{\citenamefont {Blanc}\ \emph {et~al.}(2013)\citenamefont {Blanc},
  \citenamefont {Jean}, \citenamefont {Krasheninnikov}, \citenamefont
  {Renaud},\ and\ \citenamefont {Coraux}}]{Blanc2013}%
  \BibitemOpen
  \bibfield  {author} {\bibinfo {author} {\bibfnamefont {N.}~\bibnamefont
  {Blanc}}, \bibinfo {author} {\bibfnamefont {F.}~\bibnamefont {Jean}},
  \bibinfo {author} {\bibfnamefont {A.~V.}\ \bibnamefont {Krasheninnikov}},
  \bibinfo {author} {\bibfnamefont {G.}~\bibnamefont {Renaud}}, \ and\ \bibinfo
  {author} {\bibfnamefont {J.}~\bibnamefont {Coraux}},\ }\href {\doibase
  10.1103/PhysRevLett.111.085501} {\bibfield  {journal} {\bibinfo  {journal}
  {Phys. Rev. Lett.}\ }\textbf {\bibinfo {volume} {111}},\ \bibinfo {pages}
  {085501} (\bibinfo {year} {2013})}\BibitemShut {NoStop}%
\bibitem [{\citenamefont {Baba}\ \emph {et~al.}(1993)\citenamefont {Baba},
  \citenamefont {Yamamoto},\ and\ \citenamefont {Sasaki}}]{Baba1993}%
  \BibitemOpen
  \bibfield  {author} {\bibinfo {author} {\bibfnamefont {Y.}~\bibnamefont
  {Baba}}, \bibinfo {author} {\bibfnamefont {H.}~\bibnamefont {Yamamoto}}, \
  and\ \bibinfo {author} {\bibfnamefont {T.~A.}\ \bibnamefont {Sasaki}},\
  }\href {\doibase 10.1016/0039-6028(93)91077-3} {\bibfield  {journal}
  {\bibinfo  {journal} {Surf. Sci.}\ }\textbf {\bibinfo {volume} {287-288}},\
  \bibinfo {pages} {806} (\bibinfo {year} {1993})}\BibitemShut {NoStop}%
\bibitem [{\citenamefont {Behm}\ \emph {et~al.}(1986)\citenamefont {Behm},
  \citenamefont {Brundle},\ and\ \citenamefont {Wandelt}}]{Behm1986}%
  \BibitemOpen
  \bibfield  {author} {\bibinfo {author} {\bibfnamefont {R.~J.}\ \bibnamefont
  {Behm}}, \bibinfo {author} {\bibfnamefont {C.~R.}\ \bibnamefont {Brundle}}, \
  and\ \bibinfo {author} {\bibfnamefont {K.}~\bibnamefont {Wandelt}},\ }\href
  {\doibase 10.1063/1.451299} {\bibfield  {journal} {\bibinfo  {journal} {J.
  Chem. Phys.}\ }\textbf {\bibinfo {volume} {85}},\ \bibinfo {pages} {1061}
  (\bibinfo {year} {1986})}\BibitemShut {NoStop}%
\bibitem [{Note1()}]{Note1}%
  \BibitemOpen
  \bibinfo {note} {For Gr/Ir(111) the Xe\protect \tmspace +\thinmuskip
  {.1667em}3d$_{5/2}$ peak energy is between 669.0\protect \tmspace
  +\thinmuskip {.1667em}eV and 669.2\protect \tmspace +\thinmuskip {.1667em}eV
  at all temperatures, while for bare Ir(111) at 300\protect \tmspace
  +\thinmuskip {.1667em}K it is 669.7\protect \tmspace +\thinmuskip
  {.1667em}eV. This variation is presumably due to a better extra-atomic
  screening of the implanted Xe as opposed to the trapped one \cite {Baba1993},
  but its explanation is beyond the scope of the present
  investigation.}\BibitemShut {Stop}%
\bibitem [{\citenamefont {Widdra}\ \emph {et~al.}(1998)\citenamefont {Widdra},
  \citenamefont {Trischberger}, \citenamefont {Frie\ss}, \citenamefont
  {Menzel}, \citenamefont {Payne},\ and\ \citenamefont {Kreuzer}}]{Widdra1998}%
  \BibitemOpen
  \bibfield  {author} {\bibinfo {author} {\bibfnamefont {W.}~\bibnamefont
  {Widdra}}, \bibinfo {author} {\bibfnamefont {P.}~\bibnamefont
  {Trischberger}}, \bibinfo {author} {\bibfnamefont {W.}~\bibnamefont
  {Frie\ss}}, \bibinfo {author} {\bibfnamefont {D.}~\bibnamefont {Menzel}},
  \bibinfo {author} {\bibfnamefont {S.~H.}\ \bibnamefont {Payne}}, \ and\
  \bibinfo {author} {\bibfnamefont {H.~J.}\ \bibnamefont {Kreuzer}},\ }\href
  {\doibase 10.1103/PhysRevB.57.4111} {\bibfield  {journal} {\bibinfo
  {journal} {Phys. Rev. B}\ }\textbf {\bibinfo {volume} {57}},\ \bibinfo
  {pages} {4111} (\bibinfo {year} {1998})}\BibitemShut {NoStop}%
\bibitem [{\citenamefont {Donnelly}(1978)}]{Donnelly1978}%
  \BibitemOpen
  \bibfield  {author} {\bibinfo {author} {\bibfnamefont {S.~E.}\ \bibnamefont
  {Donnelly}},\ }\href {\doibase 10.1016/S0042-207X(78)80418-7} {\bibfield
  {journal} {\bibinfo  {journal} {Vacuum}\ }\textbf {\bibinfo {volume} {28}},\
  \bibinfo {pages} {163} (\bibinfo {year} {1978})}\BibitemShut {NoStop}%
\bibitem [{\citenamefont {Lahrood}\ \emph {et~al.}(2011)\citenamefont
  {Lahrood}, \citenamefont {de~los Arcos}, \citenamefont {Prenzel},
  \citenamefont {von Keudell},\ and\ \citenamefont {Winter}}]{Lahrood2011}%
  \BibitemOpen
  \bibfield  {author} {\bibinfo {author} {\bibfnamefont {A.~R.}\ \bibnamefont
  {Lahrood}}, \bibinfo {author} {\bibfnamefont {T.}~\bibnamefont {de~los
  Arcos}}, \bibinfo {author} {\bibfnamefont {M.}~\bibnamefont {Prenzel}},
  \bibinfo {author} {\bibfnamefont {A.}~\bibnamefont {von Keudell}}, \ and\
  \bibinfo {author} {\bibfnamefont {J.}~\bibnamefont {Winter}},\ }\href
  {\doibase 10.1016/j.tsf.2011.07.040} {\bibfield  {journal} {\bibinfo
  {journal} {Thin Solid Films}\ }\textbf {\bibinfo {volume} {520}},\ \bibinfo
  {pages} {1625} (\bibinfo {year} {2011})}\BibitemShut {NoStop}%
\bibitem [{\citenamefont {Kern}\ \emph {et~al.}(1988)\citenamefont {Kern},
  \citenamefont {David}, \citenamefont {Zeppenfeld},\ and\ \citenamefont
  {Comsa}}]{Kern1988}%
  \BibitemOpen
  \bibfield  {author} {\bibinfo {author} {\bibfnamefont {K.}~\bibnamefont
  {Kern}}, \bibinfo {author} {\bibfnamefont {R.}~\bibnamefont {David}},
  \bibinfo {author} {\bibfnamefont {P.}~\bibnamefont {Zeppenfeld}}, \ and\
  \bibinfo {author} {\bibfnamefont {G.}~\bibnamefont {Comsa}},\ }\href
  {\doibase 10.1016/0039-6028(88)90347-0} {\bibfield  {journal} {\bibinfo
  {journal} {Surf. Sci.}\ }\textbf {\bibinfo {volume} {195}},\ \bibinfo {pages}
  {353} (\bibinfo {year} {1988})}\BibitemShut {NoStop}%
\bibitem [{\citenamefont {{Da Silva}}\ and\ \citenamefont
  {Stampfl}(2007)}]{DaSilva2007}%
  \BibitemOpen
  \bibfield  {author} {\bibinfo {author} {\bibfnamefont {J.~L.~F.}\
  \bibnamefont {{Da Silva}}}\ and\ \bibinfo {author} {\bibfnamefont
  {C.}~\bibnamefont {Stampfl}},\ }\href {\doibase 10.1103/PhysRevB.76.085301}
  {\bibfield  {journal} {\bibinfo  {journal} {Phys. Rev. B}\ }\textbf {\bibinfo
  {volume} {76}},\ \bibinfo {pages} {085301} (\bibinfo {year}
  {2007})}\BibitemShut {NoStop}%
\bibitem [{\citenamefont {Sheng}\ \emph {et~al.}(2010)\citenamefont {Sheng},
  \citenamefont {Ono},\ and\ \citenamefont {Taketsugu}}]{Sheng2010}%
  \BibitemOpen
  \bibfield  {author} {\bibinfo {author} {\bibfnamefont {L.}~\bibnamefont
  {Sheng}}, \bibinfo {author} {\bibfnamefont {Y.}~\bibnamefont {Ono}}, \ and\
  \bibinfo {author} {\bibfnamefont {T.}~\bibnamefont {Taketsugu}},\ }\href
  {\doibase 10.1021/jp907861c} {\bibfield  {journal} {\bibinfo  {journal} {J.
  Phys. Chem. C}\ }\textbf {\bibinfo {volume} {114}},\ \bibinfo {pages} {3544}
  (\bibinfo {year} {2010})}\BibitemShut {NoStop}%
\bibitem [{\citenamefont {Silvestrelli}\ and\ \citenamefont
  {Ambrosetti}(2015)}]{Silvestrelli2015}%
  \BibitemOpen
  \bibfield  {author} {\bibinfo {author} {\bibfnamefont {P.~L.}\ \bibnamefont
  {Silvestrelli}}\ and\ \bibinfo {author} {\bibfnamefont {A.}~\bibnamefont
  {Ambrosetti}},\ }\href {\doibase 10.1103/PhysRevB.91.195405} {\bibfield
  {journal} {\bibinfo  {journal} {Phys. Rev. B}\ }\textbf {\bibinfo {volume}
  {91}},\ \bibinfo {pages} {195405} (\bibinfo {year} {2015})}\BibitemShut
  {NoStop}%
\bibitem [{\citenamefont {He}\ \emph {et~al.}(2014)\citenamefont {He},
  \citenamefont {Robertson}, \citenamefont {Lee}, \citenamefont {Yoon},
  \citenamefont {Lee},\ and\ \citenamefont {Warner}}]{He2014}%
  \BibitemOpen
  \bibfield  {author} {\bibinfo {author} {\bibfnamefont {K.}~\bibnamefont
  {He}}, \bibinfo {author} {\bibfnamefont {A.~W.}\ \bibnamefont {Robertson}},
  \bibinfo {author} {\bibfnamefont {S.}~\bibnamefont {Lee}}, \bibinfo {author}
  {\bibfnamefont {E.}~\bibnamefont {Yoon}}, \bibinfo {author} {\bibfnamefont
  {G.}~\bibnamefont {Lee}}, \ and\ \bibinfo {author} {\bibfnamefont {J.~H.}\
  \bibnamefont {Warner}},\ }\href {\doibase 10.1021/nn504471m} {\bibfield
  {journal} {\bibinfo  {journal} {ACS Nano}\ }\textbf {\bibinfo {volume} {8}},\
  \bibinfo {pages} {12272} (\bibinfo {year} {2014})}\BibitemShut {NoStop}%
\bibitem [{\citenamefont {Michely}\ and\ \citenamefont
  {Comsa}(1991)}]{Michely1991}%
  \BibitemOpen
  \bibfield  {author} {\bibinfo {author} {\bibfnamefont {T.}~\bibnamefont
  {Michely}}\ and\ \bibinfo {author} {\bibfnamefont {G.}~\bibnamefont
  {Comsa}},\ }\href {\doibase 10.1016/0039-6028(91)90865-P} {\bibfield
  {journal} {\bibinfo  {journal} {Surf. Sci.}\ }\textbf {\bibinfo {volume}
  {256}},\ \bibinfo {pages} {217} (\bibinfo {year} {1991})}\BibitemShut
  {NoStop}%
\bibitem [{\citenamefont {Busse}\ \emph {et~al.}(2011)\citenamefont {Busse},
  \citenamefont {Lazi\'{c}}, \citenamefont {Djemour}, \citenamefont {Coraux},
  \citenamefont {Gerber}, \citenamefont {Atodiresei}, \citenamefont {Caciuc},
  \citenamefont {Brako}, \citenamefont {N'Diaye}, \citenamefont {Bl\"{u}gel},
  \citenamefont {Zegenhagen},\ and\ \citenamefont {Michely}}]{Busse2011}%
  \BibitemOpen
  \bibfield  {author} {\bibinfo {author} {\bibfnamefont {C.}~\bibnamefont
  {Busse}}, \bibinfo {author} {\bibfnamefont {P.}~\bibnamefont {Lazi\'{c}}},
  \bibinfo {author} {\bibfnamefont {R.}~\bibnamefont {Djemour}}, \bibinfo
  {author} {\bibfnamefont {J.}~\bibnamefont {Coraux}}, \bibinfo {author}
  {\bibfnamefont {T.}~\bibnamefont {Gerber}}, \bibinfo {author} {\bibfnamefont
  {N.}~\bibnamefont {Atodiresei}}, \bibinfo {author} {\bibfnamefont
  {V.}~\bibnamefont {Caciuc}}, \bibinfo {author} {\bibfnamefont
  {R.}~\bibnamefont {Brako}}, \bibinfo {author} {\bibfnamefont {A.~T.}\
  \bibnamefont {N'Diaye}}, \bibinfo {author} {\bibfnamefont {S.}~\bibnamefont
  {Bl\"{u}gel}}, \bibinfo {author} {\bibfnamefont {J.}~\bibnamefont
  {Zegenhagen}}, \ and\ \bibinfo {author} {\bibfnamefont {T.}~\bibnamefont
  {Michely}},\ }\href {\doibase 10.1103/PhysRevLett.107.036101} {\bibfield
  {journal} {\bibinfo  {journal} {Phys. Rev. Lett.}\ }\textbf {\bibinfo
  {volume} {107}},\ \bibinfo {pages} {036101} (\bibinfo {year}
  {2011})}\BibitemShut {NoStop}%
\bibitem [{Note2()}]{Note2}%
  \BibitemOpen
  \bibinfo {note} {To first approximation trapping efficiencies are obtained by
  dividing the trapping yields specified in reference \protect \citenum
  {Herbig2014} by a factor of 7.3.}\BibitemShut {Stop}%
\bibitem [{\citenamefont {Krasheninnikov}\ \emph {et~al.}(2014)\citenamefont
  {Krasheninnikov}, \citenamefont {Berseneva}, \citenamefont {Kvashnin},
  \citenamefont {Enkovaara}, \citenamefont {Bj\"{o}rkman}, \citenamefont
  {Sorokin}, \citenamefont {Shtansky}, \citenamefont {Nieminen},\ and\
  \citenamefont {Golberg}}]{Krasheninnikov2014}%
  \BibitemOpen
  \bibfield  {author} {\bibinfo {author} {\bibfnamefont {A.~V.}\ \bibnamefont
  {Krasheninnikov}}, \bibinfo {author} {\bibfnamefont {N.}~\bibnamefont
  {Berseneva}}, \bibinfo {author} {\bibfnamefont {D.~G.}\ \bibnamefont
  {Kvashnin}}, \bibinfo {author} {\bibfnamefont {J.}~\bibnamefont {Enkovaara}},
  \bibinfo {author} {\bibfnamefont {T.}~\bibnamefont {Bj\"{o}rkman}}, \bibinfo
  {author} {\bibfnamefont {P.}~\bibnamefont {Sorokin}}, \bibinfo {author}
  {\bibfnamefont {D.}~\bibnamefont {Shtansky}}, \bibinfo {author}
  {\bibfnamefont {R.~M.}\ \bibnamefont {Nieminen}}, \ and\ \bibinfo {author}
  {\bibfnamefont {D.}~\bibnamefont {Golberg}},\ }\href {\doibase
  10.1021/jp509505j} {\bibfield  {journal} {\bibinfo  {journal} {J. Phys. Chem.
  C}\ }\textbf {\bibinfo {volume} {118}},\ \bibinfo {pages} {26894} (\bibinfo
  {year} {2014})}\BibitemShut {NoStop}%
\end{thebibliography}%

\end{document}